\documentclass[fleqn,usenatbib]{mnras}

\usepackage{newtxtext,newtxmath}

\usepackage[T1]{fontenc}

\DeclareRobustCommand{\VAN}[3]{#2}
\let\VANthebibliography\thebibliography
\def\thebibliography{\DeclareRobustCommand{\VAN}[3]{##3}\VANthebibliography}

\usepackage{graphicx}	
\usepackage{amsmath}	
\usepackage{gensymb}
\usepackage{pdflscape}
\usepackage{booktabs}
\usepackage{multirow}
\usepackage{upgreek}
\usepackage{ragged2e}
\usepackage{enumitem}

\expandafter\def\expandafter\normalsize\expandafter{%
    \normalsize%
    \setlength\abovedisplayskip{3pt}%
    \setlength\belowdisplayskip{3pt}%
    \setlength\abovedisplayshortskip{3pt}%
    \setlength\belowdisplayshortskip{3pt}%
}

\bibpunct[:]{(}{)}{;}{a}{,}{,}

\title[The organic molecular content in L1517B]{The complex organic molecular content in the L1517B starless core}

\author[A. Meg\'ias et al.]{
Andr\'es Meg\'ias,$^{1}$\thanks{E-mail: \texttt{\,amegias\,@\,cab.inta-csic.es}}
Izaskun Jim\'enez-Serra,$^{1}$
Jes\'us Mart\'in-Pintado,$^{1}$
Anton I. Vasyunin,$^{2}$
\newauthor \,
Silvia Spezzano,$^{3}$
Paola Caselli,$^{3}$
Giuliana Cosentino,$^{4}$
and Serena Viti$^{\,5,6}$
\\
\\
$^{1}$ Centro de Astrobiolog\'ia (CAB), CSIC-INTA, Carretera de Ajalvir, km 4, 28850, Torrej\'on de Ardoz, Spain\\
$^{2}$ Ural Federal University, Kuybysheva st. 48, 620002, Ekaterinburg, Russian Federation\\
$^{3}$ Max Planck Institute for Extraterrestrial Physics, Giessenbachstrasse 1, 85748, Garching, Germany\\
$^{4}$ Chalmers University of Technology, SE41296, Gothenburg, Sweden\\
$^{5}$ Leiden Observatory, Leiden University, PO Box 9513, NL-2300 RA, Leiden, the Netherlands\\
$^{6}$ Department of Physics and Astronomy, University College London, Gower Street, London, WC1E 6BT, United Kingdom
}

\date{Accepted 2022 November 22. Received 2022 November 15; in original form 2022 August 30. DOI: \texttt{https://doi.org/10.1093/mnras/stac3449}.}

\pubyear{2023}

\begin{document}
\label{firstpage}
\pagerange{\pageref{firstpage}--\pageref{lastpage}}
\maketitle

\begin{abstract}
Recent observations of the pre-stellar core L1544 and the younger starless core L1498 have revealed that complex organic molecules (COMs) are enhanced in the gas phase toward their outer and intermediate-density shells. Our goal is to determine the level of chemical complexity toward the starless core L1517B, which seems younger than L1498, and compare it with the other two previously studied cores to see if there is a chemical evolution within the cores. We have carried out 3 mm high-sensitivity observations toward two positions in the L1517B starless core: the core's centre and the position where the methanol emission peaks (at a distance of $\sim$5000 au from the core's centre). Our observations reveal that a lower number of COMs and COM precursors are detected in L1517B with respect to L1498 and L1544, and also show lower abundances. Besides methanol, we only detected CH$_3$O, H$_2$CCO, CH$_3$CHO, CH$_3$CN, CH$_3$NC, HCCCN, and HCCNC. Their measured abundances are $\sim$3 times larger toward the methanol peak than toward the core's centre, mimicking the behaviour found toward the more evolved cores L1544 and L1498. We propose that the differences in the chemical complexity observed between the three studied starless cores are a consequence of their evolution, with L1517B being the less evolved one, followed by L1498 and L1544. Chemical complexity in these cores seems to increase over time, with N-bearing molecules forming first and O-bearing COMs forming at a later stage as a result of the catastrophic depletion of CO.
\end{abstract}




\section{Introduction} \label{section-intro}

In Astrochemistry, complex organic molecules (or COMs) are defined as carbon-bearing compounds with at least 6 atoms in their structure \citep {herbst-2009}. It was initially believed that COMs could only form on dust grains in the presence of a source of heat via hydrogenation, atom addition and radical-radical reactions \citep{watanabe-2002, garrod-2008}. Indeed, COMs were firstly detected in relatively hot sources with $T \gtrsim$ 100 K, such as massive hot cores \citep{hollis-2000, hollis-2006, belloche-2008, belloche-2013} or low-mass warm cores (hot corinos; \citealp{bottinelli-2004, jorgensen-2013}). However, the detection of several COMs in the gas phase in starless/pre-stellar cores and dark cloud cores (e.g., \citealp{marcelino-2007, oberg-2010, bacmann-2012, cernicharo-2012, vastel-2014, jimenez-serra-2016, jimenez-serra-2021, scibelli-2020}) indicates that there must be another chemical pathway that allows the presence of COMs in the gas phase at temperatures as low as 10 K. In the last years, there have been different proposals to explain the formation of these COMs under these conditions. However, the chemical pathways giving rise to COMs in cold cores is not well understood yet, and more observations are needed to constrain the proposed models \citep{rawlings-2013, vasyunin-2013, balucani-2015, vasyunin-2017, holdship-2019, jin-2020, punanova-2022}. For example, \citet{scibelli-2021} studied the starless core L1521E, claiming that COMs are not only formed by gas-phase reactions, bus also by surface reactions on dust grains. Other authors suggest that cosmic rays induce desorption from icy mantles on dust grains \citep{sipila-2021, redaelli-2021}. It is remarkable that methanol (CH$_3$OH), which plays a central role in the formation for larger COMs, is detected systematically in starless cores \citep{scibelli-2020}, and thought to form on the surface of dust grains, partially returning into the gas phase upon reactive desorption \citep{garrod-2006, garrod-2007, vasyunin-2017}.

\begin{figure}
\label{figure-core-image}
\begin{center}
\includegraphics[trim={0 0 0 0}, clip, width=\hsize]{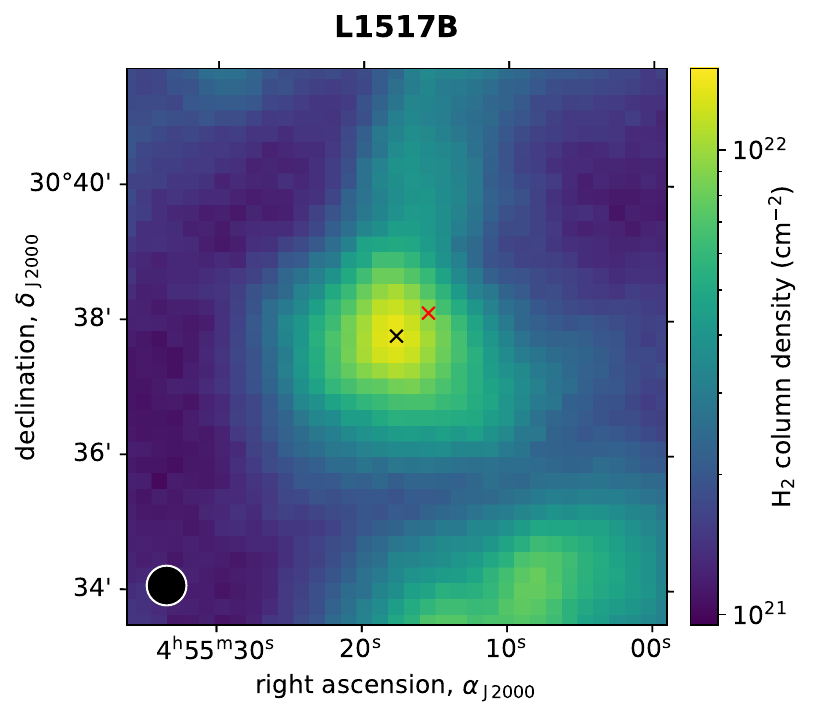}
\end{center}
\caption{$\mathrm{H}_{2}$ column density map obtained from \textit{Herschel}/\textsc{Spire} data at 0.25, 0.35 and 0.50 mm, using the same procedure as the one used for L1544 in \citet{spezzano-2016}. Crosses indicate the positions observed in the core: the dust peak (in black) and the methanol peak (in red). The beam size for each filter is 17.9, 24.2 and 35.4 arcseconds (0.25, 0.35 and 0.50 mm, respectively), although the images of the two first filters were smoothed to the resolution of 0.50 mm, which is marked with a circle at the bottom-left corner of the image. The positions of the dust and methanol peaks are obtained from \citet{tafalla-2004a, tafalla-2006}. The beam sizes were retrieved from the \textsc{Spire} Handbook: \url{https://www.cosmos.esa.int/web/herschel/spire-overview}.}
\end{figure}

In order to understand how COMs form under the cold conditions of pre-stellar/starless cores, \citet{jimenez-serra-2016, jimenez-serra-2021} investigated the radial distribution of large COMs as a function of radius in the L1544 and L1498 starless cores. Methanol tends to show a ring-like morphology circumventing the dust continuum emission (see \citealp{tafalla-2006, bizzocchi-2014, spezzano-2016, punanova-2022}). Since mapping the emission of larger COMs in starless cores requires large amounts of telescope time, Jimenez-Serra et al. (2016, 2021) observed two positions within these cores: the centre, defined by the position of the dust peak, and the location where methanol peaks. The latter position is representative of an outer, intermediate-density shell located at radii between $\sim$4000 and $\sim$11000 au from the core centre in L1544 and L1498, respectively. Several COM precursors have been found toward both cores such as  tricarbon monoxide (CCCO) and cyanoacetylene (HCCCN), but more complex molecules like acetaldehyde (CH$_3$CHO), methyl formate (CH$_3$OCHO) or dimethylether (CH$_3$OCH$_3$) have only been detected toward L1544, which is at a more advanced stage of evolution than L1498 \cite[see][]{jimenez-serra-2016, jimenez-serra-2021}. However, N-bearing COMs are detected toward L1498 with abundances close to those measured toward L1544.

In this work, we present a similar study carried out toward another starless core, L1517B, believed to be at an even earlier stage of evolution than L1498 based on the deuterium fractionation and the absence of infall motions \citep{crapsi-2005}. The goal is to compare chemical complexity measured toward L1517B with that observed in L1544 and L1498, and to establish if the observed trend of increasing chemical complexity is due to evolution \cite[see][]{jimenez-serra-2016, jimenez-serra-2021}. These cores, located in the Taurus molecular cloud complex, show different observational signatures that provide information about their stage of gravitational collapse. L1544 is classified as a pre-stellar core because it shows clear evidence of gravitational collapse toward its innermost regions as probed by the emission of N$_2$H$^+$ \citep{caselli-2002, redaelli-2019}. Its high deuterium fractionation ($N_\mathrm{N_2D^+} / N_\mathrm{N_2H^+} = 0.23 \pm 0.04$), CO depletion factor \citep[$f_\mathrm{CO} = 14 \pm 3$;][]{crapsi-2005, redaelli-2019}, and central H$_2$ density \citep[$n_\mathrm{H_2}\sim10^7$ cm$^{-3}$;][]{caselli-2022}, are also consistent with this idea. L1498 presents signatures of infall motions in the outer envelope of the core as revealed by the asymmetries observed in the line profiles of CS \citep{tafalla-2004a} but its deuterium fractionation, CO depletion factor and central H$_2$ density are lower than those measured toward L1544 ($N_\mathrm{N_2D^+} / N_\mathrm{N_2H^+} = 0.04 \pm 0.01$, $f_\mathrm{CO} = 7.5 \pm 2.5$, and $n_\mathrm{H_2}\simeq9.4 \times 10^4$ cm$^{-3}$; \citealp{crapsi-2005, tafalla-2004a}). For L1517B, the CO depletion factor and deuterium fractionation are similar to those of L1498, $f_\mathrm{CO} = 9.5 \pm 2.8$ and $N_\mathrm{N_2D^+} / N_\mathrm{N_2H^+}$ = 0.06 $\pm$ 0.01 \citep{crapsi-2005}. However, although L1517B shows a slightly higher central H$_2$ density ($n_\mathrm{H_2}\simeq2.2 \times 10^5$ cm$^{-3}$), the absence of infall motions either toward the innermost regions or toward the envelope \citep{tafalla-2004a}, suggests that L1517B is at a younger evolutionary stage than L1498 and L1544. Hence, L1517B, located at a distance of 159 pc \citep{galli-2019}, would be the dynamically youngest of the three cores.

\begin{table}
\label{table-rms-noises}
\centering
\caption{Frequency ranges, velocity resolution and RMS noise level of our observations.}
\begin{tabular}{cccc}
\hline 
\textbf{Frequency} & \textbf{Resolution} & \multicolumn{2}{c}{\textbf{RMS noise} (mK)}\tabularnewline
(GHz) & (km s$^{-1}$) & Dust peak & Methanol peak\tabularnewline
\hline 
78.2--80.0 & 0.18 & 4.8 & 3.9\tabularnewline
81.4--83.3 & 0.18 & 4.2 & 3.5\tabularnewline
83.3--85.2 & 0.17 & 2.8 & 3.3\tabularnewline
86.7--88.5 & 0.17 & 3.0 & 3.5\tabularnewline
93.5--95.7 & 0.15 & 4.1 & 3.4\tabularnewline
95.8--96.8 & 0.15 & 10.8 & 10.9\tabularnewline
97.1--99.0 & 0.15 & 4.1 & 3.6\tabularnewline
99.1--100.9 & 0.15 & 3.0 & 3.5\tabularnewline
102.4--104.2 & 0.14 & 4.1 & 4.7\tabularnewline
109.2--111.0 & 0.13 & 10.1 & 11.0\tabularnewline
\hline 
\end{tabular}
\end{table}

\begin{figure*}
\centering
\includegraphics[trim={0 7 0 0}, clip, width=\hsize]{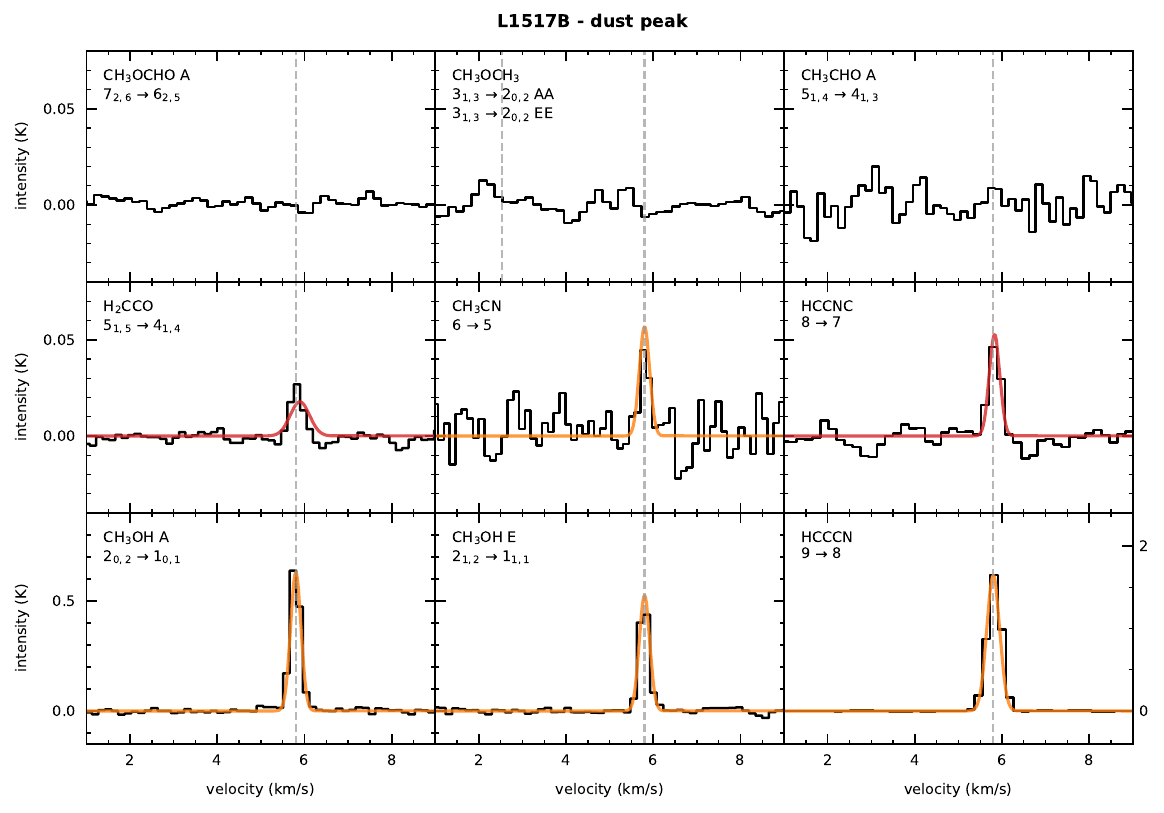}
\caption{Sample of the COM and COM precursor lines observed toward L1517B's dust continuum peak and their corresponding fits obtained with \textsc{Madcuba} (LTE, in red) and \textsc{Radex} (non-LTE, in orange). See Sections$\,$\ref{section-observations} and \ref{section-results-transitions} for more details.}
\label{figure-centre-lines}
\end{figure*}

The paper is organised as follows. In Section \ref{section-observations}, we describe the observations carried out toward L1517B. In Section \ref{section-results} we present the results of the analysis of the COMs and COM precursor emission and report the values of the derived excitation temperatures, column densities and molecular abundances. Section \ref{section-comparison} compares the abundances obtained toward L1517B with those measured toward L1498 and L1544. We also compare our results with those reported by \citet{nagy-2019} and \citet{scibelli-2021} toward the young L1521E starless core. This core has been found to be rich in COM emission despite its youth. In Section \ref{section-modelling}, we present the modelling of the COMs and COM precursor chemistry of the L1517B core, and compare the model predictions with the observations. In Section \ref{section-ratio}, we compare the column densities ratios between the N-bearing species HC$_3$N and CH$_3$CN derived toward L1517B, L1498, L1544, and L1521E, with those obtained in protostellar systems, protoplanetary discs, and comets, and discuss the observed discrepancies. Finally, in Section \ref{section-conclusions}, we summarise our conclusions.

\section{Observations} \label{section-observations}

The observations of the L1517B starless core were carried out  from September $30^\mathrm{th}$ to October $4^\mathrm{th}$ of 2020, with the Instituto de Radioastronom\'ia Milim\'etrica (IRAM) 30$\,$m telescope (Granada, Spain). As for L1544 and L1498, we observed two positions within the L1517B core: the location of the dust peak and the position where the emission of methanol peaks. These two positions have the equatorial coordinates (in J2000 system) $\alpha$ = $4^\mathrm{h}55^\mathrm{m}17\fs6$, $\delta$ = $30\degree37'44''$ for the dust continuum peak, and $\alpha$ = $4^\mathrm{h}55^\mathrm{m}15\fs7$, $\delta$ = $30\degree38'04''$ for the position of the methanol peak \citep[see][]{tafalla-2004a, tafalla-2006}. The latter is located $\sim$32$''$ away from the dust peak (see Fig. 1), which corresponds to $\sim$5000 au at a distance of 159 pc.

The high-sensitivity 3 mm spectra were obtained in frequency-switching mode using a frequency throw of 7.14 MHz. The EMIR E090 receivers were tuned at 84.37 GHz and 94.82 GHz with rejections of $\geq$10 dB. The observed frequency ranges are shown in Table \ref{table-rms-noises}. To identify possible weak spurious features in the observed spectra, we carried out part of the observations by shifting slightly the central frequencies by $\pm$20 MHz (see also \citealp{jimenez-serra-2016}). We used the narrow mode of the FTS spectrometer that provided a spectral resolution of 49 kHz, equivalent to 0.13--0.18 km s$^{-1}$ at 3 mm. Typical system temperatures ranged between 75--110 K and the telescope beam size was 22--31$\,''$ between 78 and 111 GHz; as the beams are almost gaussian and the dust and methanol peaks of L1517B are located $\sim$32$''$ away ($> 2\,\upsigma$), the contamination between both positions should be negligible (less than 5 percent). The spectra were calibrated in units of antenna temperature, $T_\mathrm{A}^*$, and converted into main beam temperature, $T_\mathrm{mb}$, by using beam efficiencies of 0.81 at 79--101 GHz and of 0.78 at 102--111 GHz.\footnote{\url{https://publicwiki.iram.es/Iram30mEfficiencies/}} The root mean square (RMS) noise level of the original observations ranged between 3 and 11 mK for both observing positions (see Table \ref{table-rms-noises}), having similar values for each frequency range than the ones obtained by \cite{jimenez-serra-2016, jimenez-serra-2021} for L1544 and L1498.


\begin{figure*}
\centering
\includegraphics[trim={0 7 0 0}, clip, width=\hsize]{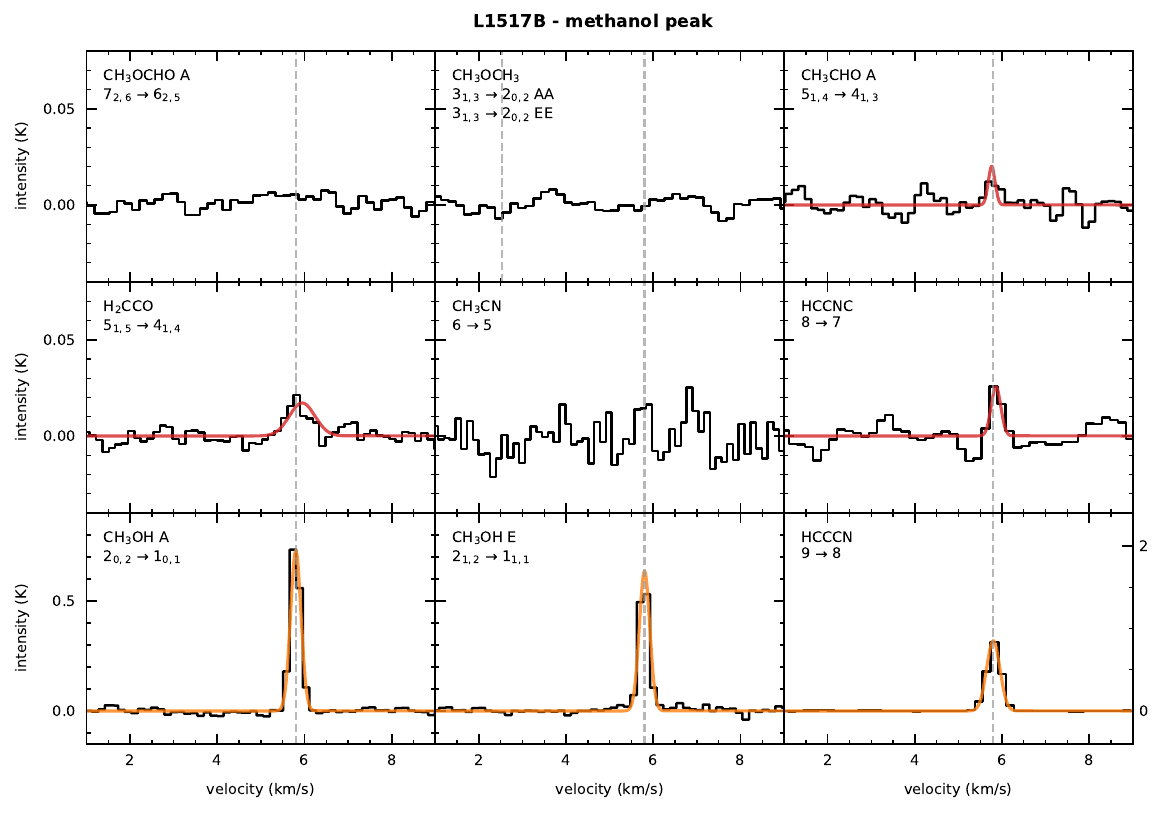}
\caption{Sample of the COM lines observed toward L1517B's methanol peak and their corresponding fits obtained with \textsc{Madcuba} (LTE, in red) and \textsc{Radex} (non-LTE, in orange). See Sections$\,$\ref{section-observations} and \ref{section-results-transitions} for more details. Acetaldehyde (CH$_3$CHO) transitions are quite noisy, but we could actually detect it because we observed several ones. It seems that acetonitrile (CH$_3$CN) may show a transition, but we could not properly fit the two covered transitions neither with \textsc{Madcuba} nor \textsc{Radex}; plus, the signal to noise is quite low (below 2 for the shown transition and below 1 for the other one).}
\label{figure-methanol-lines}
\end{figure*}

Our observations have covered the transitions of both O-bearing and N-bearing COMs and COM precursors, summarised in Table \textcolor{blue}{2}. For O-bearing species, we have targeted methanol (CH$_3$OH), methoxy (CH$_3$O), tricarbon monoxide (CCCO), ketene (H$_2$CCO), formic acid (t-HCOOH), acetaldehyde (CH$_3$CHO), methyl formate (CH$_3$OCHO), dimethylether (CH$_3$OCH$_3$), cyclopropenone (c-C$_3$H$_2$O), and propynal (HCCCHO). As N-bearing COMs and precursors, we have observed cyanoacetylene (HCCCN), isocyanoacetilene (HCCNC), vinyl cyanide (CH$_2$CHCN), acetonitrile (CH$_3$CN), and methyl isocyanide (CH$_3$NC).

The raw spectra have been analysed and reduced with \textsc{Class}, from the package \textsc{Gildas},\footnote{\url{https://www.iram.fr/IRAMFR/GILDAS/}} as well as with a Python pipeline written specifically for this purpose.\footnote{\url{https://github.com/andresmegias/gildas-class-pipeline/}} This pipeline fits baselines to the spectra using an iterative method that first masks the strongest lines using sigma-clips and then applies rolling medians and rolling averages, interpolating the masked regions with third order splines.  Finally, we used the software \textsc{Madcuba} \citep{martin-2019} to search for the molecular transitions of all targeted species, and to carry out the fitting of the molecular line profiles under the assumption of local thermodynamic equilibrium (LTE). The physical parameters derived in the fitting are the molecular column density ($N_\mathrm{obs}$), excitation temperature ($T_\mathrm{ex}$), linewidth ($\Delta v$) and LSR radial velocity ($v_\mathrm{LSR}$). For this, we used the tool \textsc{Slim} (Spectral Line Identification and Modelling) of \textsc{Madcuba}, employing the Cologne Database for Molecular Spectroscopy (CDMS; \citealp{endres-2016}) and the Jet Propulsion Laboratory (JPL) molecular catalogue \citep{pickett-1998}. In addition, for the cases of CH$_3$OH, CH$_3$CN, and HCCCN, we used the non-LTE code \textsc{Radex} \citep{vandertak-2007} to do the fitting of the lines.

\section{Results} \label{section-results}

\subsection{Detected transitions} \label{section-results-transitions}

Figs. \ref{figure-centre-lines} and \ref{figure-methanol-lines} show the observed spectra of some representative transitions of the COM and COM precursors  detected toward L1517B, while Table \textcolor{blue}{2} lists all the transitions covered in our observations with their derived line parameters. The targeted transitions are the same as in \citet{jimenez-serra-2016} and \citet{jimenez-serra-2021}, and correspond to those expected to be the brightest for an excitation temperature of $\sim$10 K.

Besides methanol, our spectra reveal the detection of other COMs and COM precursor species in L1517B: CH$_3$O, H$_2$CCO, CH$_3$CHO, HCCCN, HCCNC, CH$_3$CN, and CH$_3$NC, although acetaldehyde (CH$_3$CHO) was only detected toward the methanol peak and acetonitrile (CH$_3$CN) and methyl isocyanide (CH$_3$NC) were only detected toward the dust peak. More complex molecules such as CH$_3$OCH$_3$ or CH$_3$OCHO were not detected within our noise levels. All detected lines lie above the $3\, \upsigma$ level in integrated intensity (area), where $1\, \upsigma$ is calculated as $\Delta T (\Delta v \thinspace \updelta v)^{1/2}$, where $\Delta T$ is the RMS noise level, $\Delta v$ is the line width, and $\updelta v$ is the velocity resolution of the spectrum (see Table 1). We are confident about the detection of the transitions since their derived radial velocities correspond to the $v_\mathrm{LSR}$ of the source ($\sim$5.8 km s$^{-1}$). Moreover, except for CH$_3$OH A, CH$_3$NC and HCCNC, we have measured at least two transitions above the 3 $\upsigma$ level in integrated intensity for the detected species, stressing the identification of the species since the linewidths are narrow ($\sim\,$0.3--0.5 km s$^{-1}$). The level of line confusion at the targeted RMS noise level is also very low, as commonly found in starless/pre-stellar cores. For the non-detections, we provide upper limits to the integrated intensities by using 3 $\Delta T (\Delta v \thinspace \updelta v)^{1/2}$.

\begin{landscape}

\begin{table}
\label{table-transitions}
\caption{Most of the COMs and COM precursors transitions covered in our L1517B observations
and their derived line parameters.}

\begin{center}

\resizebox{1.06\textwidth}{!}{%

\begin{tabular}{ccccccccccc}
\hline 
 &  &  & \multicolumn{4}{c}{\textbf{Dust peak}} & \multicolumn{4}{c}{\textbf{Methanol peak}}\tabularnewline
\hline 
\textbf{Species}  & \textbf{Line }  & \textbf{Frequency }  & \textbf{Area} $^{\mathrm{a}}$  & \textbf{Linewidth}  & \textbf{LSR velocity} $^{\mathrm{b}}$  & \textbf{S/N} $^{\mathrm{c}}$  & \textbf{Area} $^{\mathrm{a}}$  & \textbf{Linewidth}  & \textbf{LSR velocity} $^{\mathrm{b}}$  & \textbf{S/N} $^{\mathrm{c}}$\tabularnewline
 &  & (MHz)  & (mK km s$^{-1}$)  & (km s$^{-1}$)  & (km s$^{-1}$)  &  & (mK km s$^{-1}$)  & (km s$^{-1}$)  & (km s$^{-1}$)  & \tabularnewline
\hline 
CH$_{3}$OH  & $2_{0,2}\rightarrow1_{0,1}$ A  & 96741.371  & 206.1 $\pm$ 1.9  & 0.278 $\pm$ 0.004  & 5.785 $\pm$ 0.002  & 109  & 241.1 $\pm$ 2.0  & 0.277 $\pm$ 0.005  & 5.788 $\pm$ 0.002  & 121 \tabularnewline
 & $2_{1,2}\rightarrow1_{1,1}$ A  & 95914.310  & < 7  & ...  & ...  & ...  & < 7  & ...  & ...  & ... \tabularnewline
 & $2_{1,1}\rightarrow1_{1,0}$ A  & 97582.798  & < 2.4  & ...  & ...  & ...  & < 2.4  & ...  & ...  & ... \tabularnewline
 & $2_{1,2}\rightarrow1_{1,1}$ E  & 96739.358  & 150.4 $\pm$ 1.9  & 0.262 $\pm$ 0.008  & 5.786 $\pm$ 0.003  & 80  & 181.6 $\pm$ 2.0  & 0.262 $\pm$ 0.003  & 5.785 $\pm$ 0.003  & 92\tabularnewline
 & $2_{0,2}\rightarrow1_{0,1}$ E  & 96744.545  & 9.7 $\pm$ 1.9  & 0.244 $\pm$ 0.012  & 5.84 $\pm$ 0.04  & 5.1  & 11.2 $\pm$ 2.0  & 0.23 $\pm$ 0.08  & 5.78 $\pm$ 0.03  & 5.6\tabularnewline
 & $2_{1,1}\rightarrow1_{1,0}$ E  & 96755.501  & < 7  & ...  & ...  & ...  & < 7  & ...  & ...  & ... \tabularnewline
\hline 
CH$_{3}$O $^{\mathrm{d}}$  & $F = 1 \rightarrow 0,\, \Lambda = -1$  & 82455.980  & $<4$  & ...  & ...  & ...  & $<3$  & ...  & ...  & ... \tabularnewline
 & $F = 2 \rightarrow 1,\, \Lambda = -1$  & 82458.252  & $5.4\pm1.2$  & $0.41\pm0.08$  & $5.81\pm0.03$  & 4.5  & $4.2\pm0.9$  & $0.39\pm0.08$  & $5.82\pm0.03$  & 4.7 \tabularnewline
 & $F = 2 \rightarrow 1,\,  \Lambda = +1$  & 82471.825  & $5.4\pm1.2$  & $0.41\pm0.08$  & $5.81\pm0.03$  & 4.5  & $4.2\pm0.9$  & $0.39\pm0.08$  & $5.82\pm0.03$  & 4.7 \tabularnewline
 & $F = 1 \rightarrow 0,\, \Lambda = +1$  & 82524.180  & $<4$  & ...  & ...  & ...  & $<3$  & ...  & ...  & ... \tabularnewline
\hline 
CCCO  & $10_{}\rightarrow9_{}$  & 96214.813  & $<7$  & ...  & ...  & ...  & $<7$  & ...  & ...  & ... \tabularnewline
\hline 
t-HCOOH  & $1_{1,1}\rightarrow0_{0,0}$  & 87926.863  & $<1.9$  & ...  & ...  & ...  & $<2.2$  & ...  & ...  & ... \tabularnewline
\hline 
H$_{2}$CCO  & $4_{1,3}\rightarrow3_{1,2}$ o- & 81586.299  & $11.1\pm1.3$  & $0.53\pm0.10$  & $5.89\pm0.04$  & 8.5  & $12.6\pm1.2$  & $0.70\pm0.25$  & $5.95\pm0.11$  & 10 \tabularnewline
 & $5_{1,5}\rightarrow4_{1,4}$ o- & 100094.510  & $9.9\pm0.9$  & $0.53\pm0.10$  & $5.89\pm0.04$  & 11  & $12.6\pm1.2$  & $0.70\pm0.25$  & $5.95\pm0.11$  & 10 \tabularnewline
 & $4_{0,4}\rightarrow3_{0,3}$ p- & 80832.189  & $21\pm4$  & $0.53\pm0.10$  & $5.89\pm0.04$  & 5.2  & $17\pm4$  & $0.70\pm0.25$  & $5.95\pm0.11$  & 4.2 \tabularnewline
\hline 
CH$_{3}$OCHO  & $7_{2,6}\rightarrow6_{2,5}$ A & 84454.754  & $<2.0$  & ...  & ...  & ...  & $<2.5$  & ...  & ...  & ... \tabularnewline
\hline 
CH$_{3}$OCH$_{3}$  & $3_{1,3}\rightarrow2_{0,2}$ EE & 82650.180  & $<3$  & ...  & ...  & ...  & $<2.4$  & ...  & ...  & ... \tabularnewline
 & $4_{1,4}\rightarrow3_{0,3}$ EE & 99326.000  & $<2.0$  & ...  & ...  & ...  & $<2.2$  & ...  & ...  & ... \tabularnewline
\hline 
CH$_{3}$CHO  & $5_{1,4}\rightarrow4_{1,3}$ A & 98900.944  & $<3$  & ...  & ...  & ...  & $4.0\pm0.8$  & $0.21\pm0.05$  & $5.765\pm0.018$  & 5.0 \tabularnewline
 & $5_{0,5}\rightarrow4_{0,4}$ A & 95963.459  & $<7$  & ...  & ...  & ...  & $<6$  & ...  & ...  & ... \tabularnewline
 & $4_{1,3}\rightarrow3_{1,2}$ A & 79150.166  & $<3$  & ...  & ...  & ...  & $3.9\pm0.7$  & $0.21\pm0.05$  & $5.765\pm0.018$  & 5.6 \tabularnewline
 & $4_{1,3}\rightarrow3_{1,2}$ E & 79099.313  & $<3$  & ...  & ...  & ...  & $3.9\pm0.7$  & $0.21\pm0.05$  & $5.765\pm0.018$  & 5.6 \tabularnewline
\hline 
c-C$_{3}$H$_{2}$O  & $6_{1,6}\rightarrow5_{1,5}$  & 79483.519  & $<3$  & ...  & ...  & ...  & $<3$  & ...  & ...  & ... \tabularnewline
\hline 
HCCCHO  & $9_{0,9}\rightarrow8_{0,8}$  & 83775.816  & $<2.0$  & ...  & ...  & ...  & $<2.5$  & ...  & ...  & ... \tabularnewline
\hline 
CH$_{3}$CN  & $6_{0}\rightarrow5_{0}$  & 110383.500  & 6.4 $\pm$ 1.9  & 0.13 $\pm$ 0.13  & 5.902 $\pm$ 0.023  & 3.4  & < 6  & ...  & ...  & ... \tabularnewline
 & $6_{1}\rightarrow5_{1}$  & 110381.372  & 11.8 $\pm$ 1.9  & 0.24 $\pm$ 0.08  & 5.81 $\pm$ 0.03  & 6.2  & < 6  & ...  & ...  & ... \tabularnewline
 & $6_{2}\rightarrow5_{2}$  & 110374.989  & < 6  & ...  & ...  & ...  & < 6  & ...  & ...  & ... \tabularnewline
\hline 
CH$_{3}$NC  & $5_{0}\rightarrow4_{0}$  & 100526.541  & $3.6\pm0.7$  & $0.4\pm0.3$  & $5.87\pm0.13$  & 5.1  & $<2.2$  & ...  & ...  & ... \tabularnewline
\hline 
CH$_{2}$CHCN  & $9_{0,9}\rightarrow8_{0,8}$  & 84945.988  & $<2.0$  & ...  & ...  & ...  & $<2.5$  & ...  & ...  & ... \tabularnewline
 & $9_{1,9}\rightarrow8_{1,8}$  & 83207.496  & $<3$  & ...  & ...  & ...  & $<2.4$  & ...  & ...  & ... \tabularnewline
 & $10_{0,10}\rightarrow9_{0,9}$  & 94276.625  & $<3$  & ...  & ...  & ...  & $<2.4$  & ...  & ...  & ... \tabularnewline
\hline 
HCCCN  & $9_{}\rightarrow8_{}$  & 81881.461  & 693.6 $\pm$ 1.0  & 0.395 $\pm$ 0.001  & 5.824 $\pm$ 0.001  & 694  & 346.7 $\pm$ 0.9  & 0.383 $\pm$ 0.002  & 5.816 $\pm$ 0.001  & 385 \tabularnewline
 & $11_{}\rightarrow10_{}$  & 100076.392  & 286.5 $\pm$ 0.7  & 0.318 $\pm$ 0.001  & 5.842 $\pm$ 0.001  & 409  & 96.3 $\pm$ 0.8  & 0.290 $\pm$ 0.003  & 5.837 $\pm$ 0.001  & 120\tabularnewline
\hline 
HCCNC  & $8_{8}\rightarrow7_{8}$  & 79484.131  & $16.1\pm1.1$  & $0.31\pm0.15$  & $5.83\pm0.07$  & 15  & $6.9\pm0.9$  & $0.29\pm0.16$  & $5.86\pm0.06$  & 7.7 \tabularnewline
 & $10_{10}\rightarrow9_{10}$  & 99354.250  & $6.8\pm0.7$  & $0.31\pm0.15$  & $5.83\pm0.07$  & 9.7  & $<2.3$  & ...  & ...  & ... \tabularnewline
\hline 
\end{tabular}

}
\end{center}

\footnotesize{Line profiles were fitted using \textsc{Madcuba}, except for methanol (CH$_3$OH), cyanoacetilene (HCCCN) and acetonitrile (CH$_3$CN), where we used \textsc{Class} (see Section \ref{section-results-column-densities} for details). ($^{\mathrm{a}}$) Uncertainties in the line area are calculated as $\Delta T (\Delta \nu \, \updelta v)^{1/2}$, with $\Delta T$ the RMS noise nevel, $\Delta v$ the line width, and $\updelta v$ the velocity resolution of the spectrum. Similarly, upper limits are calculated as 3 $\Delta T (\Delta v \, \updelta v)^{1/2}$. ($^{\mathrm{b}}$) LSR stands for \emph{local standard of rest}. ($^{\mathrm{c}}$) This refers to the signal to noise ratio in integrated intensity area. If a certain value has no uncertainty, this means that it had to be fixed so that \textsc{Madcuba} could fit the LTE model. ($^{\mathrm{d}}$) Hyperfine components of the $N = 1 \rightarrow 0,\,K=0,\, J = 3/2 \rightarrow 1/2$ transition of CH$_3$O.}

\end{table}

\end{landscape}

From Figs. \ref{figure-centre-lines} and \ref{figure-methanol-lines}, we find that the line intensities vary for each position: for the N-bearing species (CH$_3$CN, HCCNC, and HCCCN), the emission is brighter toward the core's centre with respect to the methanol peak. On the contrary, for O-bearing species (CH$_3$OH and H$_2$CCO) we find that the emission level is similar for both positions. Note that in the cases of CH$_3$OH, CH$_3$CN and HCCCN the fits were obtained with \textsc{Radex} instead of \textsc{Madcuba}. In these cases the lines were also fitted with independent gaussians using \textsc{Class}, but they are not shown in Figs. \ref{figure-centre-lines} and \ref{figure-methanol-lines}, as they are independent fits for each transition and there are only made to obtain the parameters of the lines shown in Table \textcolor{blue}{2}.
\\

\begin{table*}
\label{table-abunds}
\caption{Excitation/kinetic temperatures (\emph{$T$}), column densities (\emph{$N_{\mathrm{obs}}$}),
and abundances (\emph{$\chi_{\mathrm{obs}}\thinspace$}) of COMs and
COM precursors toward the dust and methanol peaks in L1517B.}
\label{abunds-table}

\renewcommand*{\arraystretch}{1.4}

\resizebox{\textwidth}{!}{%

\begin{tabular}{ccccccc}
\hline 
 & \multicolumn{3}{c}{\textbf{Dust peak}} & \multicolumn{3}{c}{\textbf{Methanol peak}}\tabularnewline
\hline 
\textbf{Molecule}  & \emph{$\boldsymbol{T}\thinspace(\mathrm{K})$}  & \emph{$\boldsymbol{N}_{\mathrm{\mathbf{obs}}}\thinspace(\mathrm{cm}{}^{-2})$}  & \emph{$\boldsymbol{\chi}_{\mathbf{\mathrm{\mathbf{obs}}}}\thinspace$}  & \emph{$\boldsymbol{T}\thinspace(\mathrm{K})$}  & \emph{$\boldsymbol{N}_{\mathrm{\mathbf{obs}}}\thinspace(\mathrm{cm}{}^{-2})$}  & \emph{$\boldsymbol{\chi}_{\mathrm{\mathbf{obs}}}\thinspace$}\tabularnewline
\hline 
CH$_{3}$OH A  & $10.0$  & $(4.29\pm0.11)\cdot10^{12}$  & $1.23_{-0.16}^{+0.21}\cdot10^{-10}$  & $10.0$  & $(4.96\pm0.12)\cdot10^{12}$  & $5.2_{-0.5}^{+0.6}\cdot10^{-10}$ \tabularnewline
CH$_{3}$OH E  & $10.0$  & $4.59_{-0.25}^{+2.63}\cdot10^{12}$  & $1.5_{-0.3}^{+0.7}\cdot10^{-10}$  & $10.0$  & $(5.62\pm0.20)\cdot10^{12}$  & $5.9_{-0.6}^{+0.7}\cdot10^{-10}$ \tabularnewline
CH$_{3}$OH  & $10.0$  & $8.9_{-0.3}^{+2.6}\cdot10^{12}$  & $2.7_{-0.5}^{+0.7}\cdot10^{-10}$  & $10.0$  & $(1.058\pm0.023)\cdot10^{13}$  & $1.10_{-0.10}^{+0.12}\cdot10^{-9}$ \tabularnewline
CH$_{3}$O  & $10\pm4$  & $(2.8\pm0.9)\cdot10^{11}$  & $8.0_{-2.5}^{+2.8}\cdot10^{-12}$  & $7\pm4$  & $(1.8\pm0.6)\cdot10^{11}$  & $1.9_{-0.6}^{+0.7}\cdot10^{-11}$ \tabularnewline
CH$_{3}$OCHO  & $7.7$  & $<9\cdot10^{11}$  & $<4\cdot10^{-11}$  & $9.7$  & $<1.0\cdot10^{12}$  & $<1.5\cdot10^{-10}$ \tabularnewline
CH$_{3}$OCH$_{3}$  & $7.7$  & $<1.0\cdot10^{12}$  & $<5\cdot10^{-11}$  & $9.7$  & $<5\cdot10^{11}$  & $<7\cdot10^{-11}$ \tabularnewline
CH$_{3}$CHO  & $7.7$  & $<2.5\cdot10^{11}$  & $<1.2\cdot10^{-11}$  & $9.7$  & $(2.1\pm0.4)\cdot10^{11}$  & $2.2_{-0.4}^{+0.5}\cdot10^{-11}$ \tabularnewline
t-HCOOH  & $7.7$  & $<3\cdot10^{12}$  & $<1.5\cdot10^{-10}$  & $9.7$  & $<1.9\cdot10^{12}$  & $<3\cdot10^{-10}$ \tabularnewline
c-C$_{3}$H$_{2}$O  & $7.7$  & $<1.5\cdot10^{10}$  & $<7\cdot10^{-13}$  & $9.7$  & $<5\cdot10^{10}$  & $<7\cdot10^{-12}$ \tabularnewline
H$_{2}$CCO  & $7.7\pm1.0$  & $(8.4\pm1.3)\cdot10^{11}$  & $2.4_{-0.4}^{+0.6}\cdot10^{-11}$  & $10\pm3$  & $(7.2\pm1.9)\cdot10^{11}$  & $7.6_{-2.0}^{+2.2}\cdot10^{-11}$ \tabularnewline
CCCO  & $7.7$  & $<2.4\cdot10^{11}$  & $<1.1\cdot10^{-11}$  & $9.7$  & $<1.2\cdot10^{12}$  & $<1.8\cdot10^{-10}$ \tabularnewline
HCCCHO  & $7.7$  & $<1.6\cdot10^{11}$  & $<8\cdot10^{-12}$  & $9.7$  & $<1.9\cdot10^{11}$  & $<3\cdot10^{-11}$ \tabularnewline
HCCNC  & $6.7$  & $(2.4\pm1.0)\cdot10^{11}$  & $(7\pm3)\cdot10^{-12}$  & $5.3$  & $(1.9\pm0.8)\cdot10^{11}$  & $1.9_{-0.8}^{+0.9}\cdot10^{-11}$ \tabularnewline
CH$_{2}$CHCN  & $6.7$  & $<4\cdot10^{10}$  & $<1.8\cdot10^{-12}$  & $5.3$  & $<3\cdot10^{10}$  & $<5\cdot10^{-12}$ \tabularnewline
CH$_{3}$NC  & $6.7$  & $(1.6\pm0.9)\cdot10^{10}$  & $(5\pm3)\cdot10^{-13}$  & $5.3$  & $<2.3\cdot10^{10}$  & $<3\cdot10^{-12}$ \tabularnewline
CH$_{3}$CN  & $10.0$  & $(2.1\pm0.6)\cdot10^{11}$  & $6.0_{-1.9}^{+2.1}\cdot10^{-12}$  & $10.0$  & $<2.0\cdot10^{11}$  & $<3\cdot10^{-11}$ \tabularnewline
HCCCN  & $10.0$  & $(5.16\pm0.05)\cdot10^{12}$  & $1.48_{-0.18}^{+0.25}\cdot10^{-10}$  & $10.0$  & $3.72_{-0.13}^{+0.14}\cdot10^{12}$  & $(3.9\pm0.4)\cdot10^{-10}$ \tabularnewline
\hline 
\end{tabular}
}

\renewcommand*{\arraystretch}{1}

\begin{justify}
\footnotesize{Temperatures ($T$) refer to excitation temperatures ($T_{\mathrm{ex}}$) for all the species except for methanol (CH$_{3}$OH), cyanoacetilene (HCCCN) and acetonitrile (CH$_3$CN), where they refer to kinetic temperatures ($T_{\mathrm{kin}}$). We used \textsc{Madcuba} to derive the molecular parameters from the observations except for methanol, cyanoacetilene and acetonitrile, where we used \textsc{Radex}. Molecular abundances were calculated using an $\mathrm{H}_{2}$ column density of $(3.5 \pm 0.5) \times 10^{22}$ cm$^{-2}$ for the dust continuum peak and of $(9.6 \pm 1.0) \times 10^{21}$ cm$^{-2}$ for the position of the methanol peak. For the non-detections and also for some detections, we had to fix the excitation temperature ($T_{\mathrm{ex}}$) so that \textsc{Madcuba} could fit the column density ($N_{\mathrm{obs}}$).}
\end{justify}

\end{table*}

\subsection{Molecular column densities and excitation temperatures} \label{section-results-column-densities}

\subsubsection{LTE analysis}
\label{section-LTE}

Using the tool \textsc{Slim} from \textsc{Madcuba}, we performed the LTE fitting to the observed line profiles for each species, obtaining the parameters shown in Table \textcolor{blue}{2}. As in most cases we detected several lines of the same molecular species, this fitting allows us to obtain its column density ($N_\mathrm{obs}$) and excitation temperature ($T_\mathrm{ex}$). These parameters are also shown in Table \textcolor{blue}{3}.

However, there were cases in which the \textsc{Madcuba} fitting algorithm did not converge. This can be due because the signal-to-noise ratio is not high enough, or also because we only have one transition detected. In these cases, we had to fix one or more parameters, so that the algorithm could converge. Similarly, for non-detections we had to fix the excitation temperature so that \textsc{Madcuba} could fit an upper limit for the column density.

In general, when we needed to fix the temperature, we used the fitted value for H$_2$CCO for oxygen-bearing species, and the fitted one for HCCCN for nitrogen-bearing species, as these are molecules with several transitions which also have high signal-to-noise ratio. These values are, for H$_2$CCO, 7.7 K and 9.7 K for the dust and methanol peaks, respectively, and, for HCCCN, 6.7 K and 5.3 K. For cases where we had to fix the central velocity of the line, we used a value of $v_\mathrm{LSR}$ = 5.8 km s$^{-1}$.

For methyl isocyanide (CH$_3$NC) toward the core centre and isocyanoacetylene (HCCNC) toward the methanol peak, we only detected 1 transition, so we had to fix the excitation temperature to $T_\mathrm{ex}$ = 6.7 K for CH$_3$NC and to $T_\mathrm{ex}$ = 5.3 K for HCCNC (which are the corresponding values for HCCCN at the dust and methanol peaks, respectively).

For ketene (H$_2$CCO), we calculated the ortho-to-para ratio (we have detected two ortho transitions and one para transition), by carrying out a separate fit of the transitions and by dividing the resulting column densities of the ortho and para species. We obtain an ortho-to-para ratio of $1.0_{-0.3}^{+0.4}$ for the position of the dust peak and $0.9_{-0.4}^{+0.5}$ for the methanol peak. This ratio is different to what has been previously observed towards the molecular cloud TMC-1 and the pre-stellar core L1689B, where the ratio was $\sim$3 \citep{ohishi-1991, bacmann-2012}. An exact value of 3 would correspond to the statistical ratio due to the nuclear spin degeneracy. However, lower values are also possible for the lowest temperatures, like in the case of formaldehyde (H$_2$CO), where an ortho-to-para ratio of 1 would indicate that this molecule is formed under thermal equilibrium at a temperature of $\sim$10 K \citep{kahane-1984}. We note, however, that we have only detected two ortho lines and one para line of ketene, and therefore the derived ortho-to-para ratios can be subject to significant uncertainties.

\subsubsection{Non-LTE analysis}
\label{section-nonLTE}

For some cases, in particular for methanol (CH$_3$OH), cyanoacetilene (HCCCN) and acetonitrile (CH$_3$CN), we have decided to use the non-LTE code \textsc{Radex} \citep{vandertak-2007} to infer the column density $N_\mathrm{obs}$ of the species instead of \textsc{Madcuba}. In the case of methanol {and acetonitrile, we did it because \textsc{Madcuba} could not fit properly all the transitions, and the errors were greater than $3\, \upsigma$. For the case of cyanoacetilene (HCCCN), we opted for \textsc{Radex} because it allows us to estimate the H$_2$ number density, as we have two transitions. As input parameters for these calculations, we assumed the linewidth obtained with the \textsc{Madcuba} fit ($\sim$0.28 km s$^{-1}$) and a kinetic temperature of 10 K (as inferred for all radii across the L1517B starless core; see \citet{tafalla-2004a} and Section \ref{section-modelling} below). Apendix \ref{appendix-a} describes the procedure used to derive the phyisical properties and their uncertainties from the modelling.

In the case of CH$_3$OH, we have to distinguish between the A and E species, making a separate fit for each molecule. We used the collisional coefficients with H$_2$ given by \citet{rabli-2005}. In addition to deriving the column densities reported in Table \textcolor{blue}{3}, we also derived the H$_2$ number density from the observations of methanol E, as this parameter can be determined by the intensity ratio of the detected lines. Unfortunately, for the case of methanol A we only detected one transition so the column density was estimated by assuming the H$_2$ number density obtained by \citet{tafalla-2004a}: $2.20 \times 10^5$ cm$^{-3}$ for L1517B core centre and $1.23 \times 10^5$ cm$^{-3}$ for the location of the methanol peak. For methanol E we could derive the H$_2$ number density, although the uncertainties are high, obtaining values of $5_{-4}^{+8} \times 10^4$ cm$^{-3}$ for the core's centre and $2.0_{-0.6}^{+0.9} \times 10^5$ cm$^{-3}$ for the methanol peak. These values are consistent, within the uncertainties, with those derived by \citet{tafalla-2004a}, as we have to take into account the possible uncertainties in the determination of the density profiles from the absorption coefficient of the dust. As for the excitation temperatures, they are close to the assumed kinetic temperature of 10 K, consistent with those expected for the derived H$_2$ densities, although this is biased by the fact that the lowest temperature for which we have collisional rates is 10 K.

For CH$_3$CN, we fixed the H$_2$ number density to the values from \citet{tafalla-2004a} (as the signal-to-noise ratio was not high enough to get a good fit). For the core's centre, one of the lines could be fitted within $2\, \upsigma$, providing a better fit than the one by \textsc{Madcuba}. For the methanol peak, we did not detect any transition, so we obtained an upper limit to its column density from \textsc{Radex}.

For HCCCN, we have used the two transitions to derive the column density and the H$_2$ number density towards both positions. We used the collisonal coefficients with H$_2$ given by \citet{faure-2016}. Fig. \ref{figure-loss-min-2} shows the results obtained from the non-LTE analysis for the core's centre. As expected the derived H$_2$ densities and column densities are related. We obtained the HCCCN column densities shown in Table \textcolor{blue}{3} and H$_2$ densities of $1.80_{-0.04}^{+0.04} \times 10^5$ cm$^{-3}$ and $5.56_{-0.18}^{+0.19} \times 10^4$ cm$^{-3}$ for the dust and methanol peaks, respectively. Although these results present lower uncertainties than those of methanol, our H$_2$ densities are still consistent with the prediction by \citet{tafalla-2004a} taking into account the beam size of our observations and the uncertainties in the determination of the density from dust emission, as we previously discussed. In the following sections, we use the values from \citet{tafalla-2004a} because they are more direct than the derivation through \textsc{Radex}, and because the authors assume the same dust emissivity ($\kappa = 0.005$ cm$^2$ g$^{-1}$ for 1.2 mm) as that considered by \citet{crapsi-2005} for inferring the CO depletion factor of L1517B, which has been employed in our astrochemical simulations (Section \ref{section-modelling}).

\subsection{Molecular abundances} \label{section-results-abundances}

Once we know the column densities of COMs and COM precursors detected toward L1517B, we can compute the molecular abundances of these species by using the H$_2$ column densities measured toward the positions of the dust and methanol peaks (Table \textcolor{blue}{3}). For the position of the dust peak, the H$_2$ column density of $(3.5 \pm 0.5) \times 10^{22}$ cm$^{-2}$ was obtained using the 1.2 mm continuum emission observed by \citet{tafalla-2004a} with the \textsc{Mambo} 1 mm bolometer array at IRAM 30$\,$m telescope. For the position of the methanol peak, however, we employed the data from the \textit{Herschel} space telescope at 0.25, 0.35 and 0.5 mm assuming a dust optical depth index of $\beta = 1.5$ \citep{spezzano-2016}; the derived H$_2$ column density is $(9.6 \pm 1.0) \times 10^{21}$ cm$^{-2}$. As discussed in \citet{jimenez-serra-2021}, this is the best procedure to probe the total H$_2$ column density respectively toward the innermost regions (with bolometers at 1.2 mm) and outer shells (with \textit{Herschel}) in starless cores.

\begin{figure*}
\begin{center}
\includegraphics[trim={0 13 0 0}, clip, width=\hsize]{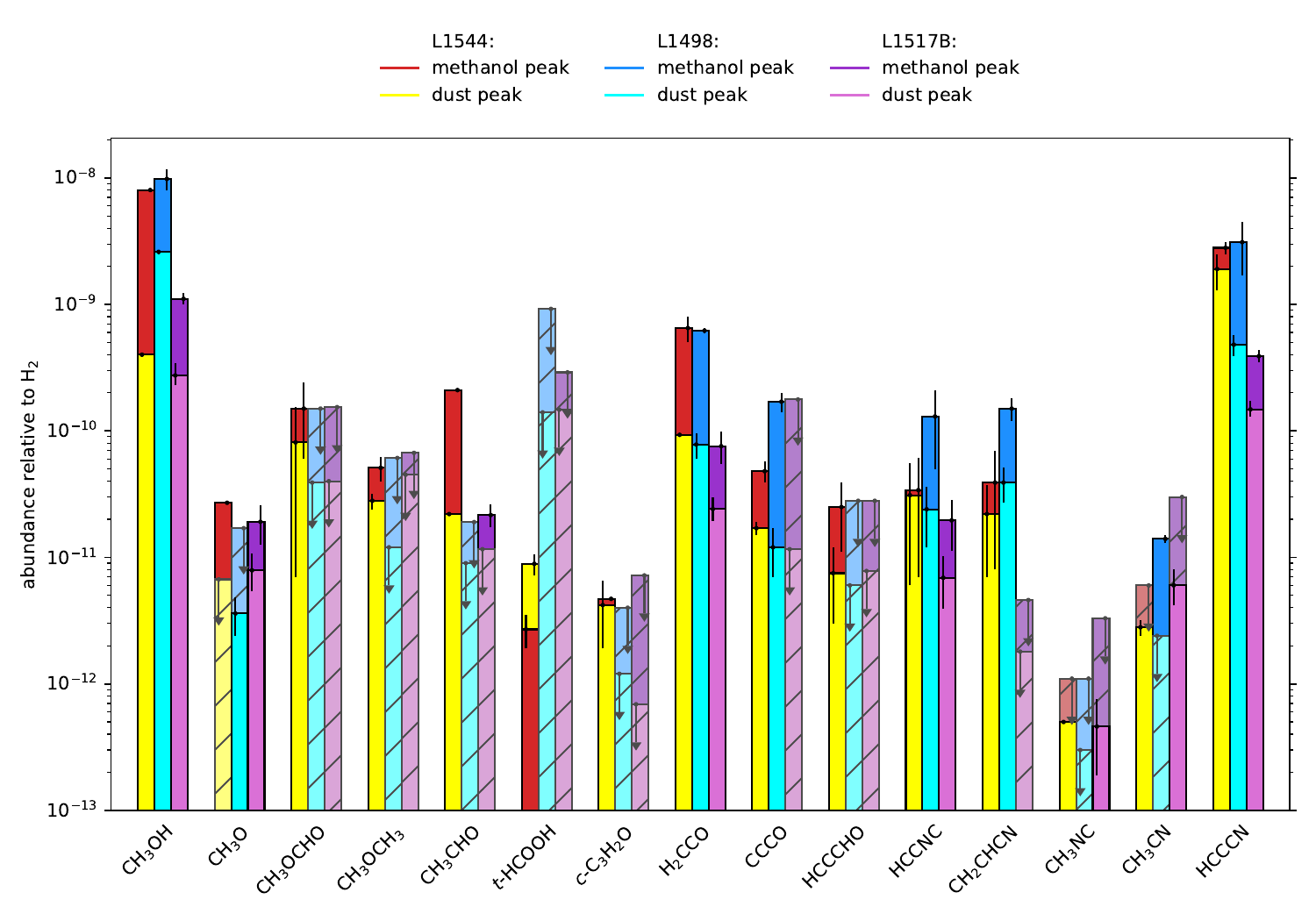}
\end{center}
\caption{Bar plot of the abundances of different COMs and COM precursors measured for the cores L1517B, L1498 and L1544, toward the dust peak (lighter colors) and the methanol peak (darker colors) for each of them. Upper limits are indicated both with arrows at the top of the corresponding bar and with stripes. Abundances for L1544 are taken from \citet{jimenez-serra-2016} and \citet{jimenez-serra-2021}, while for L1498 they are taken from \citet{jimenez-serra-2021}. Note that this is not a cumulative bar plot, so the upper edge of each bar, for both colors (that is, for both the centre and the methanol peak, for each core), indicates the abundance of the corresponding species.}
\label{figure-barplot}
\end{figure*}

From Table \textcolor{blue}{3}, we can see that the abundance of the detected species is enhanced toward the position of the methanol peak with respect to the core's centre. For methanol (CH$_3$OH), its abundance is enhanced by a factor of $\sim$4, while for the rest of detected species in both positions (CH$_3$O, H$_2$CCO, HCCNC, and HCCCN) the factor of enhancement is $\sim$3. Acetaldehyde (CH$_3$CHO) is only detected in the methanol peak, the ratio would be $\gtrsim$2. As for acetonitrile (CH$_3$CN) and methyl isocianide (CH$_3$NC) we only find lines toward the dust peak, and the ratio would be $\lesssim\,$6. Therefore, for these three molecules, the lower/upper limits for the factor of enhancement toward the methanol peak are consistent with those of the molecules detected in both positions. Additionally, the ratio of the abundances of methanol A and E species toward the dust and methanol peak is, respectively, $0.91_{-0.32}^{+0.08}$ and $0.88_{-0.04}^{+0.04}$, which is similar to what has been observed in the L1544 and L1498 cores \citep{jimenez-serra-2016, jimenez-serra-2021} and to the expected values of $\sim\,$0.7--1.0 \citep{wirstrom-2011}.

\section{Comparison with other cores} \label{section-comparison}

In this Section, we compare our results obtained for the starless core L1517B with those of the L1544 and L1498 cores \citep[see][]{jimenez-serra-2016, jimenez-serra-2021} and also with those of the L1521E core \citep[see][]{nagy-2019, scibelli-2021}.

\subsection{Comparison with L1498 and L1544}
\label{section-comparison-L1498L1544}

For L1517B, L1498 and L1544, the same molecular species have been observed toward two positions (i.e., the dust peak and the methanol peak for each of them), which allows us to obtain information about the radial distribution of the abundances of COM and COM precursors within the cores. 

Fig. \ref{figure-barplot} presents the abundances of the COM and COM precursor molecules observed toward these three cores and for the two measured positions. The abundances measured toward the dust and methanol peaks are plotted respectively in yellow and red for the L1544 pre-stellar core \citep[][]{jimenez-serra-2016}, in light and dark blue for the L1498 starless core \citep[][]{jimenez-serra-2021}, and in light and dark purple for the L1517B starless core (this work). 

Consistently with L1498 and L1544, the abundances of the species detected toward L1517B tend to be higher toward the position of the methanol peak than toward the dust peak (t-HCOOH toward L1544 is an exception). Fig. \ref{figure-barplot} also shows that the L1517B core presents a lower number of detections than L1544 and a similar number to L1498. For the N-bearing species, we have detected only HCCCN and HCCNC toward both positions of L1517B from a total of 5 targeted species, although we have found CH$_3$CN and CH$_3$NC toward the dust peak. A significant difference with respect to L1498 and L1544 is that vinyl cyanide (CH$_2$CHCN) is not detected toward L1517B, showing upper limits to its abundance that are factors $\gtrsim$10 lower than the abundances measured toward L1498 and L1544. Following the same trend, for O-bearing species only four COMs and COM precursor species (CH$_3$OH, CH$_3$O, CH$_3$CHO, and H$_2$CCO) out of 10 have been detected toward L1517B, and in the case of CH$_3$CHO it is only detected toward the methanol peak.

If we compare in detail L1517B and L1498, we can see that the first one has a level of chemical complexity somewhat lower than L1498. Firstly, the abundances of the COMs and COM precursors detected toward L1517B tend to show lower abundances than (or are comparable to) those measured toward L1498. Secondly,  
if we count the total number of detections for any position, we obtain 13 detections for L1517B and 14 for L1498, which may seem a small difference. However, if we consider the detections and non-detections for each position, the lower level of chemical complexity for the L1517B starless core becomes more apparent.

To quantitatively measure this we have calculated, for each pair of cores, the fraction of molecules with the same state of detection (detected or not detected). That is, for each source, we build a vector with one entry for each molecule and position (15 molecules $\times$ 2 positions, 30 entries), whose value is 1 if there is a detection and 0 if not. Then, we build another vector with the same number of entries, that are 1 if the two input entries (of the two cores) have the same value and 0 otherwise. Finally, we sum up all the elements of this vector and divide it by the number of entries, in order to normalise it. We call the resulting number the \textit{fractional similarity}, whose values lie between 0 and 1. In this way, if for each position and molecule both entries have the same value toward both sources (either a detection or a non-detection), the fractional similarity is 1. On the contrary, if for each position and molecule the entries are different, the resulting similarity is 0. If we compute the fractional similarity between L1517B and L1498 we obtain $0.70 \pm 0.03$.\footnote{The value used for the uncertainty is $1/30 \simeq 0.03$, which is the difference in the fractional similarity produced by one entry that is different in the two vectors.} This indicates that, although the two cores are similar in terms of number of detections, they are truly different regarding their level of chemical complexity since the similarity, clearly less than 1, is produced by 9 different entries in the vectors (that is, in the detections for each molecule and position).

For the sake of completeness, we have also computed the fractional similarity between L1544 and L1517B, which is $0.57 \pm 0.03$, and between L1544 and L1498, which is $0.43 \pm 0.03$. This indicates that the three cores show a significantly different level of chemical complexity in terms of detections and non detections, and the two more similar cores would be L1498 and L1517B.

\begin{table}
\label{table-complexity}
\caption{Weighted geometric mean of the abundance with respect to $\mathrm{H}_{2}$, $\left\langle \chi_{\mathrm{obs}}\right\rangle $,
and weighted mean molecular mass, $\left\langle m_{\mathrm{molec}}\right\rangle $ for the targeted species in L1517B, L1498, and L1544.}

\renewcommand*{\arraystretch}{1.4}

\begin{tabular}{ccccccc}
\toprule 
\multicolumn{3}{c}{\textbf{Source}} & \multicolumn{2}{c}{$\boldsymbol{\left\langle \chi_{\mathrm{obs}}\right\rangle }$ $\left(\times10^{-12}\right)$} & \multicolumn{2}{c}{$\boldsymbol{\left\langle m_{\mathrm{molec}}\right\rangle }$ $\left(\mathrm{g/mol}\right)$}\tabularnewline
\midrule 
\multirow{2}{*}{{\footnotesize{}L1517B}} & \multicolumn{2}{c}{{\scriptsize{}dust peak\enskip{}\enskip{}}} & $7.1_{-1.5}^{+1.5}$  & \multirow{2}{*}{$15_{-3}^{+3}$} & $45.5_{-0.3}^{+0.3}$  & \multirow{2}{*}{$45.61_{-0.27}^{+0.20}$}\tabularnewline
 & \multicolumn{2}{c}{{\scriptsize{}meth. peak}} & $23_{-5}^{+5}$  &  & $45.7_{-0.4}^{+0.3}$  & \tabularnewline
\multirow{2}{*}{{\footnotesize{}L1498}} & \multicolumn{2}{c}{{\scriptsize{}dust peak\enskip{}\enskip{}}} & $10.5_{-2.1}^{+2.1}$  & \multirow{2}{*}{$32_{-5}^{+5}$} & $46.18_{-0.24}^{+0.18}$  & \multirow{2}{*}{$46.25_{-0.17}^{+0.15}$}\tabularnewline
 & \multicolumn{2}{c}{{\scriptsize{}meth. peak}} & $53_{-10}^{+10}$  &  & $46.32_{-0.24}^{+0.23}$  & \tabularnewline
\multirow{2}{*}{{\footnotesize{}L1544}} & \multicolumn{2}{c}{{\scriptsize{}dust peak\enskip{}\enskip{}}} & $19_{-4}^{+4}$  & \multirow{2}{*}{$30_{-4}^{+4}$} & $46.8_{-0.4}^{+0.3}$  & \multirow{2}{*}{$46.46_{-0.21}^{+0.15}$}\tabularnewline
 & \multicolumn{2}{c}{{\scriptsize{}meth. peak}} & $42_{-7}^{+7}$  &  & $46.13_{-0.19}^{+0.16}$  & \tabularnewline
\cmidrule{2-7} \cmidrule{3-7} \cmidrule{4-7} \cmidrule{5-7} \cmidrule{6-7} \cmidrule{7-7} 
\end{tabular}

\renewcommand*{\arraystretch}{1.0}

\,

\footnotesize{The weights for the mean abundance are the molecular masses for each molecule. The weights for the mean molecular mass are the abundance for each molecule in logarithmic scale. The values on the second columns for each magnitude are the arithmetic means of the magnitudes for both positions. To deal with uncertainty propagation and upper limits we made simulations with statistical distributions (see Appendix \ref{appendix-b}).}

\end{table}

We have used two other methods to evaluate the level of chemical complexity toward the three cores. First, we have computed for each source and position the mean molecular mass of the species weighted by the abundances in a logarithmic scale.\footnote{We used logarithmic weights due to the range of the values of our abundances, which encloses several orders of magnitude.} This can be viewed as an estimate of the level of complexity of each source and position, as COMs tend to have more atoms with increasing complexity, and thus a greater molecular mass. From Table \ref{table-complexity}, we find that the lowest value of the mean molecular mass is obtained for L1517B, while the greatest one is for L1544, with L1498 being close to it. This would indicate that L1517B presents the lowest level of chemical complexity of the three cores, followed by L1498 and L1544. Additionally, we also computed for each source and position the geometric mean of the abundance of the species weighted by the molecular mass.\footnote{We used the geometric mean due to the range of the values of our abundances.} The resulting value represents a measure of the amount of the more complex species in the core, as we are giving larger weights in the mean to more complex species. As shown by Table \ref{table-complexity}, L1517B shows the lowest mean abundance, followed by L1498 and L1544, which present barely the same quantity. All of this indicates that the level of chemical complexity in the L1517B starless core seems to be lower than that of L1498, and of L1544.

From all these results, we suggest that L1517B is in a less chemically evolved stage than L1498 and L1544. This is supported by the deuterium fractionation of L1517B, similar to that of L1498, but L1517B shows no evidence of infall motions \citep{crapsi-2005, tafalla-2004a}. Additionally, L1517B shows the lowest level of chemical complexity within the three cores, presenting only the simplest COMs and COM precursors and with lower abundances in most of the cases.

If we take into account the possible time evolution between the three cores (with L1517B being the youngest and L1544 the oldest), we propose a scenario in which the N-bearing molecules would form first in starless cores. Then, as the cores evolve, they would accrete gas from the surrounding molecular cloud, becoming denser and yielding a strong depletion of carbon monoxide (CO). At the pre-stellar core stage, the catastrophic depletion of CO takes place, triggering the formation of O-bearing COMs and COM precursors as observed in L1544 \citep{vasyunin-2017}. Therefore, chemical complexity would increase over time. In the next section, we explore this scenario by modelling the chemistry of N-bearing and O-bearing COMs and COM precursors toward L1517B. 

\subsection{Comparison with L1521E}
\label{section-comparison-L1521E}

\begin{figure}
\begin{center}
\includegraphics[trim={0 13 0 0}, clip, width=\hsize]{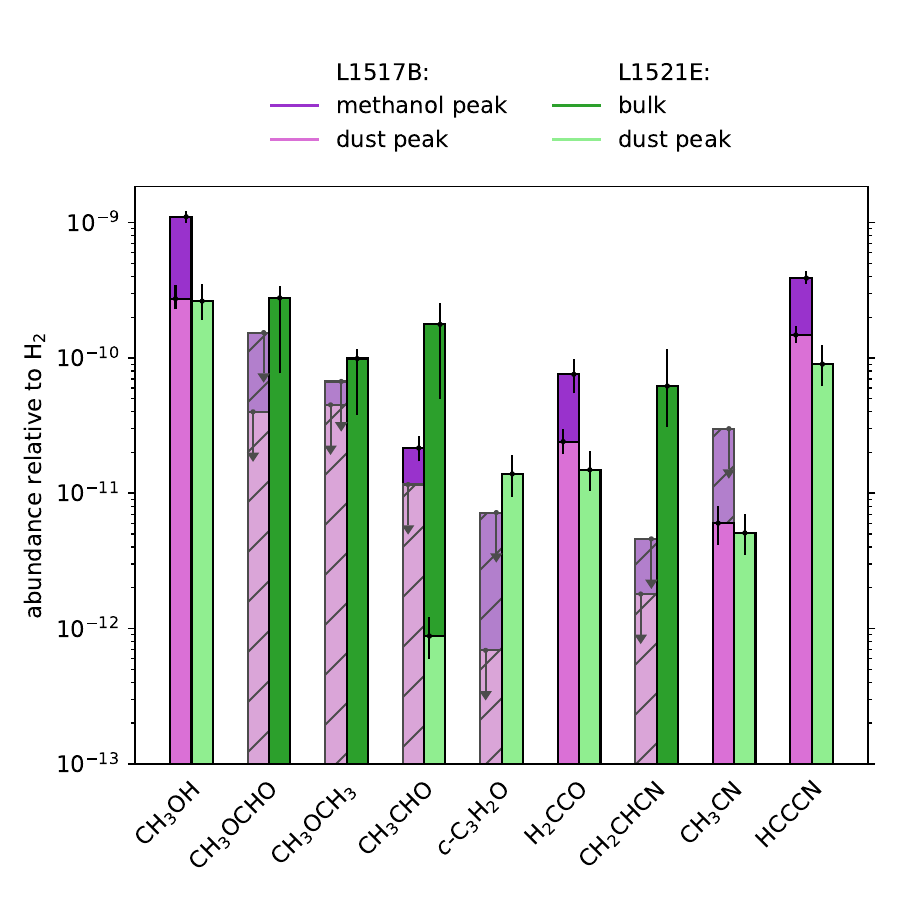}
\end{center}
\caption{Bar plot of the abundances of different COMs and COM precursors measured for the young starless cores L1517B (in light and dark purple) and L1521E (in light and dark green). For L1517B, the values are as in Figure$\,$\ref{figure-barplot}. For L1521E, the abundances shown in dark green refer to the \textit{bulk COM emission} observed by \citet{scibelli-2021} toward L1521E with a beam size of $\sim$70$''$. The large beam includes not only the dust peak position but also the methanol ring (peak located at 22--29$\,''$ from L1521E's dust peak). Abundances from L1521's dust peak are shown in light green and are taken from \citet[][]{nagy-2019}. Upper limits are indicated with both arrows and stripes.}
\label{figure-barplot-alt}
\end{figure}

L1521E is another well-studied starless core located in the Taurus molecular cloud. Like L1517B, L1521E is also considered to be young. Indeed, L1521E shows no evidence of infall motions, and has a central density of $2.7 \times 10^5$ cm$^{-3}$ \citep{tafalla-2004b}. Its CO depletion factor is $f_\mathrm{CO} = 4.3 \pm 1.6$ \citep{nagy-2019}, lower than those of L1517B and L1498. Although its deuterium fractionation $N_\mathrm{N_2D^+} / N_\mathrm{N_2H^+}$ has not been measured yet, several authors agree in the fact that this core is young (e.g., \citealp{tafalla-2004b, kong-2015, scibelli-2021}). \cite{nagy-2019} observed the molecular line emission toward the dust peak of this core (including some COMs and COM precursors) using the IRAM 30$\,$m telescope. More recently, \cite{scibelli-2021} detected four COMs toward L1521E (CH$_3$OCHO, CH$_3$OCH$_3$, CH$_3$CHO, and CH$_2$CHCN) using the 12$\,$m ARO (Arizona Radio Observatory) telescope, with a much larger beam size ($\sim$70$''$ versus 22--31$\,''$ for the IRAM 30$\,$m telescope). This implies that while the observations of \citet{nagy-2019} would cover the location of just the dust peak, the observations of \citet{scibelli-2021} would cover both the dust peak and the methanol ring, as its methanol peak is found 22--29$\,''$ away from the core center \citep{nagy-2019}. We thus refer to the observations of \citet{scibelli-2021} as those performed toward the \textit{bulk} of the core.


Among the 15 targeted species measured toward L1517B, there are 9 of them that have also been targeted toward L1521E: six toward the dust peak and four for the bulk of the core (with just one molecule in common for both positions). Fig. \ref{figure-barplot-alt} reports the abundances of the common species observed toward L1517B and L1521E. For the data under the {\it bulk} label, we use the values derived by \cite{scibelli-2021} assuming two source size cases: the lower limits of the error bars refer to the values obtained assuming a beam filling factor of 1, while the error bars upper limits refer to the abundances obtained with the best-fit source size of 35$''$. Note, however, that vinyl cyanide (CH$_2$CHCN) is an exception since its abundance was only derived assuming a beam filling factor of 1 \citep[the source size for this molecule could not be estimated;][]{scibelli-2021}.

In general, there is a good agreement between the abundances measured toward the dust peak position of the two cores (see light purple and light green bars in Fig. \ref{figure-barplot-alt}), with the sole exception of cyclopropenone (c-C$_3$H$_2$O). In contrast, the differences are bigger between the bulk abundances of L1521E and the abundances measured toward the dust and methanol peaks of L1517B. However, if we take into account the uncertainties with the error bars upper/lower limits, the abundances of L1521E's bulk are largely consistent with those measured toward both positions of L1517B, with the exception of vinyl cyanide (CH$_2$CHCN). A possible explanation for this 
would be that L1521E is more chemically evolved than L1517B, lying closer to L1498 than to L1517B. Alternatively, although these cores are located in the same molecular cloud complex, the local physical and chemical properties of their environment could differ, causing significant differences in their final chemical composition. Studies of the COM chemical content toward a larger sample of starless cores are needed to establish whether differences in the environment  affect the COM chemical evolution in starless cores. 

We finally note that calculations of the fractional similarity and of the geometric means of the abundance and of the molecular mass for L1521E, cannot be performed since the number of data available for this core is rather small.

\section{Modelling the formation of COMs and COM precursors in L1517B} \label{section-modelling} 

\subsection{The model} \label{section-modelling-model}

We have modelled the chemistry of COMs and COM precursors toward the starless core L1517B by using the 0D gas-grain chemical code \textsc{Monaco} \citep{vasyunin-2013, vasyunin-2017}, which has been successfully applied previously to the starless cores L1544 and L1498 \citep{jimenez-serra-2016, jimenez-serra-2021}. Our goal is to compare the observed molecular abundances with those predicted by the model, and to infer the radial distribution of COMs and COM precursors toward L1517B and its age.

\begin{figure}
\begin{center}
\includegraphics[trim={0 8 0 0}, clip, width=0.95\hsize]{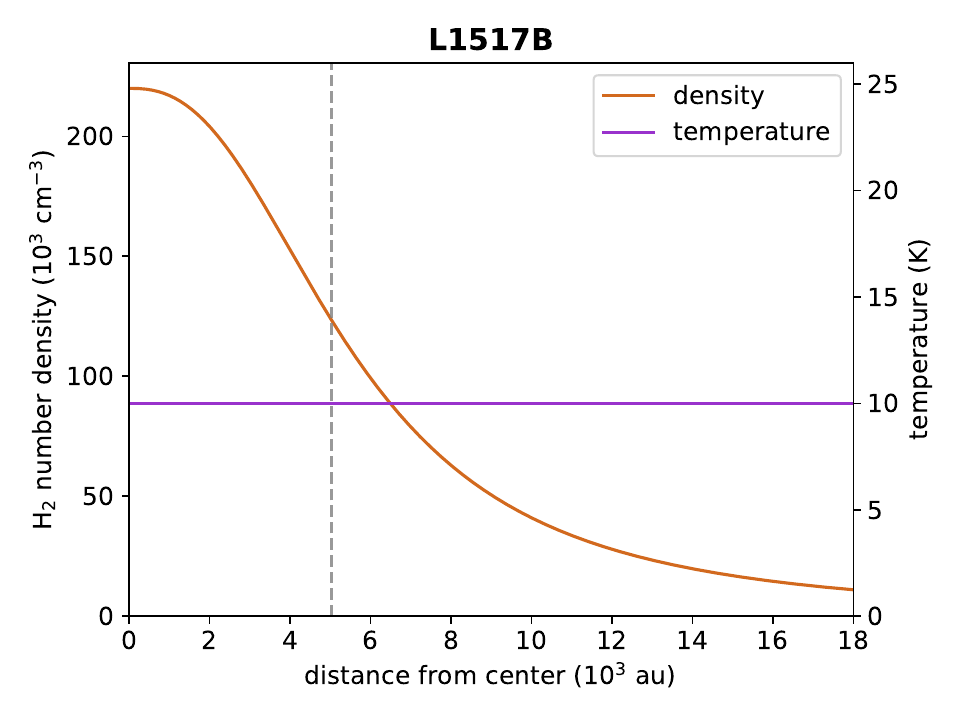}
\end{center}
\caption{H$_2$ gas density and temperature profiles used for modelling the chemistry of COMs and COM precursors toward L1517B. These radial profiles are taken from \citet{tafalla-2004a}. The vertical gray dashed line indicates the distance of the observed methanol peak with respect to the centre of the core.}
\label{figure-density-profile}
\end{figure}

\begin{figure}
\begin{center}
\includegraphics[trim={0 8 0 0}, clip, width=0.96\hsize]{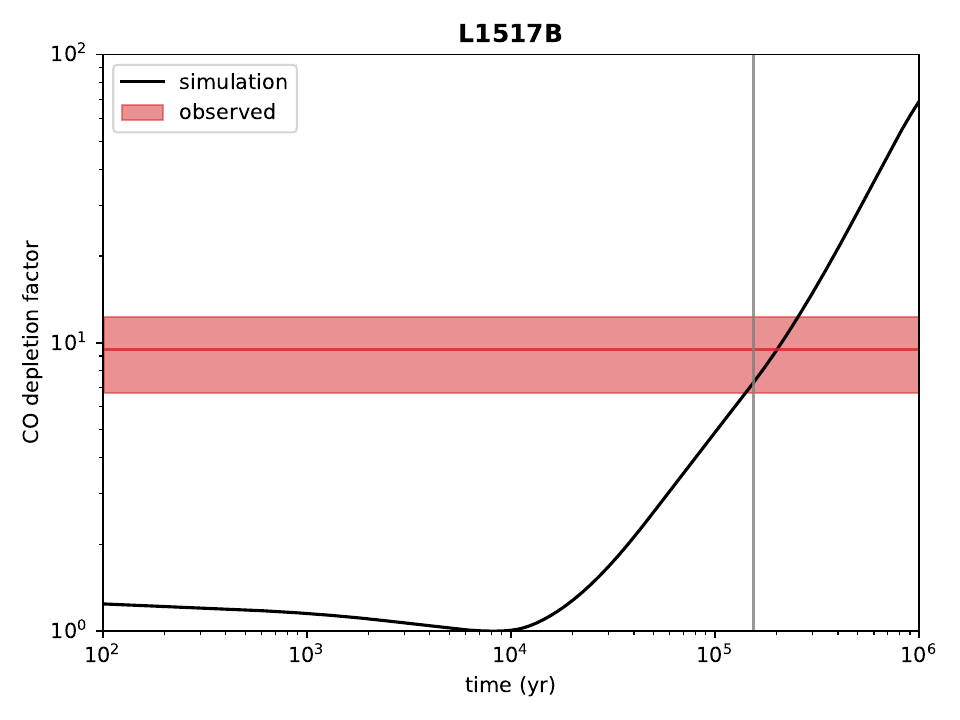}
\end{center}
\caption{Evolution of the simulated CO depletion factor over time. The blue line and shaded region show the value $f_\mathrm{CO} = 9.8 \pm 2.8$, derived by \citet{crapsi-2005}, while the gray line indicates the best-fit age for L1517B.}
\label{figure-depletion-factor}
\end{figure}

\begin{figure*}
\begin{center}
\includegraphics[trim={0 12 0 0}, clip, width=\hsize]{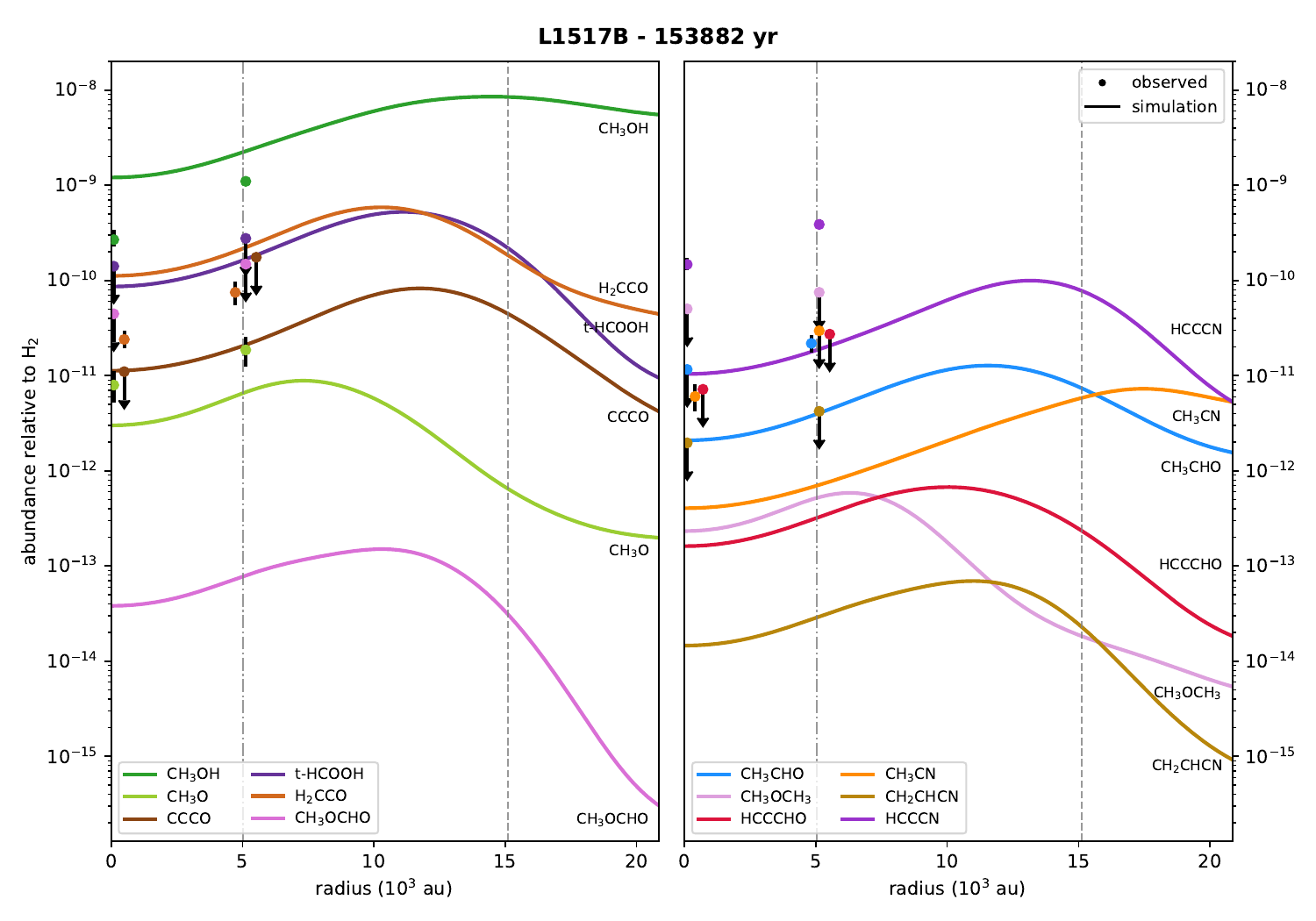}
\end{center}
\caption{Radial distribution of the modelled abundances of the COMs and COM precursors observed for L1517B. Colour curves represent the modelled abundances, while dots and vertical arrows represent the observed abundances, and their upper limits, toward the two observed positions in L1517B, the dust and methanol peaks. The position of some of the dots (observed COMs and COM precursor abundances) has been slightly shifted to enhance visibility. The dotted-dashed gray line indicates the position of the observed methanol peak, while the gray dashed line indicates the position of the methanol peak according to the chemical simulations.}
\label{figure-com-profile}
\end{figure*}

The \textsc{Monaco} chemical code is a rate equations-based, three-phase (gas, ice surface, ice bulk) numerical model that provides the evolution of the fractional abundances of atomic and molecular species with time. To obtain the radial distribution of the abundances of COMs and COM precursors in L1517B, the code is run for a grid of distances on the density and temperature profile of L1571B, assuming physical parameters to be constant in time. The physical structure of the L1517B starless core is obtained from the H$_2$ gas density profile by \citet{tafalla-2004a}, and the gas kinetic temperature distribution inferred by the same authors using NH$_3$ observations. The assumed H$_2$ gas density profile for L1517B is:
\begin{equation}
    \indent
    \label{dens-profile-eq}
    n(r) \;=\; \frac {2.2 \times 10^5 \mathrm{\;cm}^{-3}} {1 + \left( \frac {r} {35"} \right)^{2.5}}  \;\;\mathrm{,}
\end{equation}
where $r$ is the angular distance (in arcseconds). As for the temperature radial profile, we have used a constant temperature of $T$ = 10 K \citep[see][]{tafalla-2004a}. Fig. \ref{figure-density-profile} shows both profiles as a function of distance to the centre of the core. In our model, we assume that the dust and gas temperatures are equal. 

Given the low temperatures in starless cores ($T$ $\simeq$ 10 K), the precursors of COMs are formed on the surface of dust grains via hydrogenation processes. Once formed, a small fraction of these compounds are non-thermally desorbed from dust grains into the gas phase, where they undergo gas-phase chemical reactions that yield the observed COMs \citep[][]{vasyunin-2013,vasyunin-2017}. As non-thermal desorption processes, the code includes UV photo-desorption, cosmic-ray-induced desorption \citep{hasegawa-1993} and reactive (chemical) desorption \citep{minissale-2016}. The chemical network for the O-bearing and N-bearing COMs and precursors is the same as the one used for the L1498 starless core \citep[see the details in][]{jimenez-serra-2021}.  
As initial abundances, the model employs the results of a simulation of a diffuse cloud model with constant gas density of 100 cm$^{-3}$ and gas temperature of 20 K for 10$^7$ years, using the \textit{low metals} initial abundances EA1 from \citet{wakelam-2008}.

\subsection{Comparing simulations and observations} \label{section-modelling-results}

Using the initial conditions explained in Section \ref{section-modelling-model}, we have simulated the chemical evolution of L1517B by running the \textsc{Monaco} code for $10^6$ years. To constrain the age of the core, we have used the CO depletion factor of $f_\mathrm{CO}$ = 9.5 $\pm$ 2.8 measured by \citet{crapsi-2005} toward L1517B, and the location of the methanol peak at a distance of 5000 au. Interestingly, our model cannot reproduce simultaneously the location of the methanol peak at 5000 au and the CO depletion factor of $\sim\,$7--12. In Fig. \ref{figure-depletion-factor}, we show the evolution of the predicted CO depletion factor over time and its observed value. Note that the \textsc{Monaco} code considers the telescope beam size when calculating the CO depletion factor. By comparing the model results with the CO and COMs and COM precursor observations, the best agreement is reached after 1.5 $\times$ 10$^5$ years. The CO depletion factor at this time is $\sim$7.3, which is on the lower edge of the estimated interval for the CO depletion factor (see Fig. \ref{figure-depletion-factor}).

At this time of chemical evolution, however, the location of the methanol peak in the model does not match the observed one toward L1517B. This is clearly shown in Fig. \ref{figure-com-profile}, where we present the radial distribution of the abundance of COMs and COM precursors modelled for L1517B, together with the measured values and upper limits (see dots and vertical arrows). As shown in Fig. \ref{figure-com-profile}, the location of the methanol peak in the model is at $\sim$15000 au away from the dust peak (vertical dashed line), which is a factor of 3 further away that observed (at $\sim$5000$\,$au; see dashed-dotted vertical line).

The origin of this discrepancy likely arises from the fact that the central H$_2$ gas density toward L1517B is $2.2 \times 10^5$ cm$^{-3}$, i.e. significantly lower than measured toward the L1544 pre-stellar core. If this density were higher, as in L1544 ($\sim$10$^7$ cm$^{-3}$;  \citealp{caselli-2022}), the CO depletion factor (9.5 $\pm$ 2.8) would be reached earlier in the simulations, enabling a better match between the modelled and the observed location of the methanol peak at 5000 au. This would also help reconciling the age of the L1517B starless core predicted by the model (of $\sim$1.5$\times$10$^5$ years) with the one expected from observations. As already mentioned, L1517B does not show evidence for gas accretion either toward the innermost regions or from the outer envelope, as observed toward L1498 or L1544, which suggests that L1517B is at an earlier stage of evolution, or at least, at a similar evolutionary stage as L1498 (note that they both have similar central H$_2$ gas densities and deuterium fractionation values; \citealp{tafalla-2004a, crapsi-2005}). Another possible explanation of this discrepancy would be a prominent asphericity of the core, although from Fig. \ref{figure-core-image} it seems that L1517B is rather spherical.

For the abundances of COMs and COM precursors, Table \textcolor{blue}{5} compares the values between the modelled and observed abundances toward the positions of the dust and methanol peaks in L1517B. All the modelled and observed abundances agree within a factor of 10 except for CH$_3$CN in the core's center and HC$_3$N in both positions. Actually, at earlier times of the simulations, where the simulated methanol peak is closer to the observed one, abundances of those two species are closer to the observed ones. In any case, we should take into account that astrochemical models entail intrinsic uncertainties derived from uncertainties from some of the constants used in the simulation, such as the chemical reaction rates \citep{vasyunin-2004, wakelam-2005, wakelam-2010}.

Comparing the predicted age for L1517B ($1.5 \times 10^5$ yr) with those predicted for L1498 and L1544, we find that, contrary to what we thought, L1517B would not be the youngest core, as the predicted ages for L1498 and L1544 are, respectively, $9.8 \times 10^4$ yr and $1.6 \times 10^5$ yr \citep{jimenez-serra-2016, jimenez-serra-2021}.
Therefore, according to the chemical ages based on similar initial conditions, L1517B would be more evolved than L1498 and less than L1544. However, although L1517B shows some concentration of H$_2$ at its centre, there is no clear sign of current contraction motions toward this core. This, together with the observed level of chemical complexity, seems to suggest that L1517B is less evolved that L1544 and L1498. Note also that our chemical modelling does not include the dynamical evolution of the core and, thus, the chemical age derived by our model does not necessarily coincide with the actual dynamical age of the core. In fact, if the initial conditions of the cores were different, the final COM chemical composition of the cores would be significantly altered as compared to our chemical modelling. 

For L1521E, \cite{scibelli-2021} also used the \textsc{Monaco} code to estimate the age of the core (of $\sim$6 $\times 10^4$ yr). However, they could only reproduce the observed abundances of acetaldehyde (CH$_3$CHO), out of a total of five modelled COMs (with differences in the predicted abundances by factors of 20-160). Therefore, additional aspects in the chemical modelling of these young starless cores such as different initial conditions and/or dynamical evolution of the cores, may be required.

\section{Evolution of the HC$_3$N$\,/\,$CH$_3$CN abundance ratio across low-mass star formation} \label{section-ratio}


\begin{table*}
\label{table-model-observations}
\centering
\caption{Observed and modelled abundances ($\chi_{\mathrm{obs}}$, $\chi_{\mathrm{mod}}$)
for several COMs and COM precursors in L1517B.}

\renewcommand*{\arraystretch}{1.4}

\begin{tabular}{ccccccc}
\hline 
 & \multicolumn{3}{c}{\textbf{Dust peak}} & \multicolumn{3}{c}{\textbf{Methanol peak}}\tabularnewline
\cline{2-7} \cline{3-7} \cline{4-7} \cline{5-7} \cline{6-7} \cline{7-7} 
\textbf{Species} & $\boldsymbol{\chi}_{\mathrm{\mathbf{obs}}}$ & $\boldsymbol{\chi}_{\mathrm{\mathbf{mod}}}$ & \textbf{agreement} & $\boldsymbol{\chi}_{\mathrm{\mathbf{obs}}}$ & $\boldsymbol{\chi}_{\mathrm{\mathbf{mod}}}$ & \textbf{agreement}\tabularnewline
\hline 
CH$_{3}$OH & $2.7_{-0.5}^{+0.7}\times10^{-10}$ & $1.21\times10^{-9}$ & + & $1.11_{-0.10}^{+0.12}\times10^{-9}$ & $2.23\times10^{-9}$ & +\tabularnewline
CH$_{3}$O & $8.0_{-2.5}^{+2.9}\times10^{-12}$ & $3.0\times10^{-12}$ & + & $1.9_{-0.6}^{+0.7}\times10^{-11}$ & $6.6\times10^{-12}$ & +\tabularnewline
CCCO & $<\thinspace8\times10^{-12}$ & $1.12\times10^{-11}$ & + & $<\thinspace1.3\times10^{-10}$ & $2.05\times10^{-11}$ & +\tabularnewline
t-HCOOH & $<\thinspace1.0\times10^{-10}$ & $8.6\times10^{-11}$ & + & $<\thinspace2.0\times10^{-10}$ & $1.63\times10^{-10}$ & +\tabularnewline
H$_{2}$CCO & $2.4_{-0.5}^{+0.6}\times10^{-11}$ & $1.12\times10^{-10}$ & + & $7.6_{-2.0}^{+2.3}\times10^{-11}$ & $2.17\times10^{-10}$ & +\tabularnewline
CH$_{3}$OCH$_{3}$ & $<\thinspace3\times10^{-11}$ & $2.32\times10^{-13}$ & + & $<\thinspace5\times10^{-11}$ & $5.2\times10^{-13}$ & +\tabularnewline
CH$_{3}$CHO & $<\thinspace8\times10^{-12}$ & $2.09\times10^{-12}$ & + & $2.2_{-0.4}^{+0.5}\times10^{-11}$ & $4.0\times10^{-12}$ & +\tabularnewline
HCCCHO & $<\thinspace5\times10^{-12}$ & $1.62\times10^{-13}$ & + & $<\thinspace2.1\times10^{-11}$ & $3.2\times10^{-13}$ & +\tabularnewline
CH$_{3}$CN & $6.0_{-1.9}^{+2.1}\times10^{-12}$ & $4.0\times10^{-13}$ & - & $<\thinspace3\times10^{-11}$ & $7.0\times10^{-13}$ & +\tabularnewline
CH$_{2}$CHCN & $<\thinspace1.2\times10^{-12}$ & $1.46\times10^{-14}$ & + & $<\thinspace3\times10^{-12}$ & $2.9\times10^{-14}$ & +\tabularnewline
HCCCN & $1.48_{-0.18}^{+0.24}\times10^{-10}$ & $1.04\times10^{-11}$ & - & $3.9_{-0.4}^{+0.4}\times10^{-10}$ & $1.85\times10^{-11}$ & -\tabularnewline
\hline 
\end{tabular}

\renewcommand*{\arraystretch}{1}

\footnotesize{The modelled and measured abundances agree (+) or not (-) within a factor of 10.}
\end{table*}

Recent observational campaigns have been devoted to investigate the chemical composition in COMs (both O-bearing and N-bearing) towards all stages in the process of star formation. In this Section, we compare the information available for several sources for two of these COMs and COM precursors, the N-bearing species cyanoacetylene (HC$_3$N) and acetonitrile (CH$_3$CN), with the abundances obtained towards our limited sample of starless cores (L1544, L1498, L1517B, and L1521E). We have selected these two molecular species because they have been measured systematically in the past few years across a variety of objects, from Class 0/I protostars \citep{bergner-2017} to Class II protoplanetary discs (see the \textsc{Maps} ALMA large program; e.g., \citealp{ilee-2021}), and comets \citep{biver-2021}.

Fig. \ref{figure-abunds-ratio} reports the values of the column density ratio of HC$_3$N$\,/\,$CH$_3$CN for the starless cores L1544, L1498 and L1517B (for both positions measured in these cores), the starless core L1521E in the dust peak, twelve Class 0/I protostellar systems (studied by \citealp{bergner-2017}), four Class II protoplanetary discs (studied by \citealp{ilee-2021}), and two comets (studied by \citealp{biver-2021}, and references therein). To obtain this ratio, we used the column densities for both molecules, except for the comets, for which we employed the abundances of both molecules with respect to water. From Fig. \ref{figure-abunds-ratio}, it seems that there is a trend for the starless cores to present much higher HC$_3$N$\,/\,$CH$_3$CN ratios (of $\sim\,$200--700) than those measured toward Class 0/I protostars and Class II protoplanetary discs, and toward comets (in this case, their HC$_3$N$\,/\,$CH$_3$CN abundance ratios reach values even lower than 10). The core L1521E would be an exception, as its ratio is similar to those of the protostars. In any case, the observed general trend suggests that the observed HC$_3$N$\,/\,$CH$_3$CN abundance ratio decreases when starless cores enter into the Class 0 phase, remaining approximately constant until the Class II protoplanetary phase, and droping by two orders of magnitude for comets. 

The observed range of values for the HC$_3$N$\,/\,$CH$_3$CN ratio could be explained by the fact that it is difficult to form HC$_3$N at late evolutionary stages in the process of low-mass star formation. Indeed, HC$_3$N is mainly a gas-phase product whose formation is favoured by the low-densities and cold temperatures found in the outer envelopes of pre-stellar systems and protostars (see Section 5.1 in \citealp{bergner-2017}, for more details). In the same work, the authors report similar CH$_3$CN abundances for different kinds of sources (young stellar objects, comets, hot cores, and hot corinos), which would explain the decreasing trend in the HC$_3$N$\,/\,$CH$_3$CN abundance ratio with time. 

\section{Conclusions} \label{section-conclusions}

The high-sensitivity spectra obtained for the starless core L1517B reveal a chemical complexity a bit poorer than that observed in the core L1498, and quite lower than in L1544, as it only presents a few and simple O-bearing and N-bearing species, such as CH$_3$O, H$_2$CCO, CH$_3$CN, and HCCCN, and in general with lower abundances than in L1498 and L1544. Quantitatively, the geometric mean of the abundance of the targeted species in L1517B is about half than in L1498 and L1544. Similarly to these latter cores, the targeted molecules are more abundant by a factor of $\sim$3 towards the methanol peak of L1517B, located at $\sim$5000 au from the core's centre.

We have also modelled the chemical evolution of L1517B and, except for HCCCN and CH$_3$CN, our simulations agree with the observed abundances and upper limits within a factor of 10. The model also predicts the observed enhancement of the abundances as we move further away from the core's centre, although we overestimate the distance of the methanol peak by a factor of 3. Our model suggests that L1517B is chemically older than L1498, but the absence of infall motions and its relatively poor chemical complexity makes us think that this core is at an earlier evolutionary stage than L1498. Actually, our model would predict a younger age for L1517B than L1498 if the central density of the core were higher.

We propose a scenario in which N-bearing molecules are formed first, followed by O-bearing molecules once the catastrophic depletion of CO takes place. Furthermore, complexity increases with time, with bigger molecules being formed at later stages of the core, as observed in L1544. We note, however, that the starless core L1521E does not seem to follow this trend, although the sample of molecules studied is smaller than for the other three cores, and the beam of the ARO observations is larger. More observational studies of starless and pre-stellar cores are needed in order to clearly determine the influence of the environment, initial conditions and dynamics on the chemical evolution of the cores.

Finally, we have also studied the HC$_3$N$\,/\,$CH$_3$CN abundance ratio measured in sources at different stages in the formation of a low-mass stellar system, observing a decreasing trend over time, which could be explained by the adverse conditions of the late stages of the low-mass star formation to form HC$_3$N.

\begin{figure*}
\begin{center}
\includegraphics[trim={0 12 0 0}, clip, width=0.75\hsize]{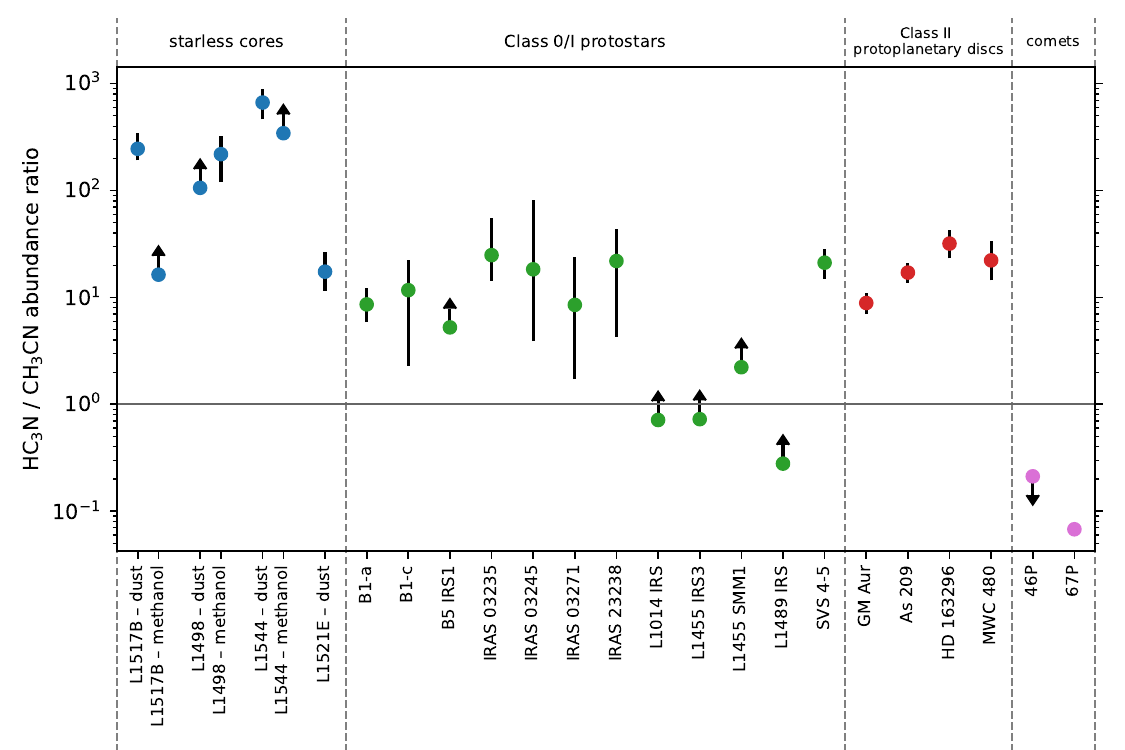}
\end{center}
\caption{Abundance ratio between cyanoacetylene (HC$_3$N) and acetonitrile (CH$_3$CN) for different sources: the starless cores L1517B, L1498 and L1544 (with both positions, the dust and methanol peaks), the starless core L1521E in the dust peak, 12 Class 0/I protostars, 4 Class II proto-planetary discs, and 2 comets (46P and 67P). The references for each source are: L1517B, this work; L1498, \citet{jimenez-serra-2021}; L1544, \citet{jimenez-serra-2016}; L1521E, \citet{nagy-2019}; Class 0/I protostellar systems, \citet{bergner-2017}; Class II protoplanetary discs, \citet{ilee-2021}; comets, \citet{biver-2021}. The points of the plot as well as the original data used to obtain the ratios can be found in Appendix \ref{appendix-c}.}
\label{figure-abunds-ratio}
\end{figure*}

\section*{Acknowledgements}
We thank an anonymous referee for their valuable comments and suggestions, which improved the quality of the paper. We also thank J. Aguirre and M. Fern\'andez-Ruz for their input about ways to measure complexity. A.M. aknowledges support from grant PRE2019-091471 under project MDM-2017-0737-19-2 funded by the Spanish Ministry of Science and Innovation / State Agency of Research, MCIN/AEI/10.13039/501100011033, and by `ESF, Investing in your future'. I.J.-S., J.M.-P. and A.M. aknowledge support from grant PID2019-105552RB-C41 by  MCIN/AEI/10.13039/501100011033. The work by A.V. is supported by the Russian Ministry of Science and Higher Education via the State Assignment contract FEUZ-2020-0038. IRAM is supported by INSU/CNRS (France), MPG (Germany), and IGN (Spain).

\section*{Data Availability}

The original spectra used in this work will be shared on reasonable request to the corresponding author. Many of the codes used to process and visualise the data can be downloaded from GitHub\footnote{\url{https://github.com/andresmegias/gildas-class-pipeline/}, \\and the same link but changing the name of the repository (\texttt{gildas-class-\\pipeline}) by \texttt{radex-line-fitter} and \texttt{richvalues}.} or from the website of our research group.\footnote{\fontsize{7.8}{8}{\url{https://sites.google.com/view/cab-ccismsf-group/software/}}}



\bibliographystyle{mnras}
\bibliography{refs}




\appendix

\section {Derivation of the methanol column density and its uncertainty with \textsc{Radex}} \label{appendix-a}

In this Section, we explain the procedure we have followed to determine the column densities of methanol, acetonitrile and cyanoacetilene, and their uncertainties, using the non-LTE molecular excitation code \textsc{Radex} \citep{vandertak-2007}. It is a radiative transfer model that allows us to predict the line intensity of the selected molecule from the following parameters:

\begin{itemize}[leftmargin=6pt]
    \item \textbf{Molecule}. It has to be included in the \textsc{Radex} molecule list, therefore having collisional cross-section derived, experimentally or theoretically. 
    \item \textbf{Spectral range}. Frequency range for the transitions to be predicted for the selected molecule.
    \item \textbf{Excitation conditions}.
    \begin{itemize}[topsep=0pt]
        \item Molecular hydrogen number density, $n_\mathrm{H_2}$.
        \item Kinetic temperature of the molecule, $T_\mathrm{kin}$.
        \item Background temperature, $T_\mathrm{bg}$.
    \end{itemize}
    \item \textbf{Radiative transfer parameters}.
    \begin{itemize}[topsep=0pt]
        \item Column density, $N$.
        \item Line width, $\Delta v$.
    \end{itemize}
\end{itemize}

For each species, the results of the calculation would be a set of intensity lines, $\{\tilde{I_i}\}$, where $i$ is the index of the line. In our case, the molecules are methanol (CH$_3$OH) A and E, acetonitrile (CH$_3$CN), and cyanoacetilene (HC$_3$N), and we have observations for their lines in a specific spectral range (see Table \textcolor{blue}{2}), so we have a set of observed intensities, $\{I_i\}$. We have an uncertainty for each line (the RMS noise), so we will also have a set of uncertainties $\{\Delta I_i\}$. We use the frequency range of the observed lines, plus a little margin, for the \textsc{Radex} calculations. The rest of the parameters except the column density (and optionally the H$_2$ number density) are known (see Section \ref{section-results-column-densities} for more details):

\begin{itemize}[leftmargin=6pt]
    \item Molecular hydrogen number density (not fixed for the case of methanol E and cyanoacetilene):
    \begin{itemize}[topsep=0pt]
        \item Core's centre: $n_\mathrm{H_2}$ = $2.20 \times 10^{5}$ cm$^{-3}$ \citep{tafalla-2004a}.
        \item Methanol peak: $n_\mathrm{H_2}$ = $1.24 \times 10^{5}$ cm$^{-3}$ \citep{tafalla-2004a}.
    \end{itemize}
    \item Background temperature: $T_\mathrm{bg}$ = 2.73 K (temperature of the cosmic microwave background).
    \item Kinetic temperature: $T_\mathrm{kin}$ = 10 K \citep{tafalla-2004a}.
    \item Line width: $\Delta v \sim 0.3$ km s$^{-1}$ (we choose for each case the result of the \textsc{Madcuba} fit).
\end{itemize}

Our goal is to optimise the column density and optionally also the H$_2$ number density (when more than one transition is available) to minimise the difference between the observed lines, $\{I_i\}$, and the predicted ones, $\{\hat{I_i}\}$, taking into account the uncertainties, $\{\Delta I_i\}$. To measure that difference, we define a loss function, $\mathcal{L}$, which we choose to be the chi-square ($\upchi^2$) error:
\begin{equation}
    \indent
    \label{equation-loss}
    \mathcal{L} \;=\; \sum_i {\left ( \frac {\hat I_i - I_i} {\Delta I_i} \right ) ^2} \,\, .
\end{equation}
In this way we are measuring the quadratic error weighted with the squared inverse of the uncertainty, so that a difference between the observations and the model contributes less to the loss if the uncertainty on the line intensity is greater.

\begin{figure}
\begin{center}
\includegraphics[trim={0 8 0 0}, clip, width=\hsize]{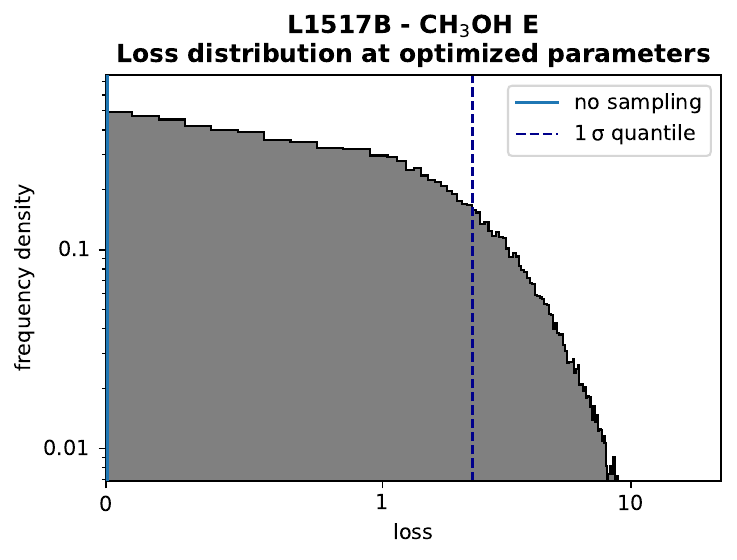}
\end{center}
\caption{Histogram of the distribution of losses obtained from the intensities of the observed lines and their uncertainties, for the case of methanol E in the dust peak of L1517B. The scale in the horizontal axis is symmetrical-logarithmic with threshold 1, that is, it is linear from 0 to 1 and logarithmic beyond 1. The blue line on the left shows the value of the loss without taking into account the uncertainties of the observed lines (Equation \ref{equation-loss}), which is almost 0; and the blue dashed line indicates the 68.27 percentile.}
\label{figure-loss-hist}
\end{figure}

Once we have properly defined our loss function, we can minimise it with respect to the column density for each of our 8 cases (four species, methanol A and E, cyanoacetilene, and acetonitrile; and two positions, dust and methanol peaks) and also with respect to the H$_2$ number density for methanol E and cyanoacetilene. We did this with Python,\footnote{\url{https://www.python.org/}} using the methods \textsc{Cobyla} and Nelder-Mead within the function \texttt{minimize} from the library SciPy.\footnote{\url{https://scipy.org/}}

Finally, we have to estimate the uncertainty for the optimised value/s of our parameter/s. To do so, we take a threshold for the loss, greater than the minimised value, which defines a lower and an upper uncertainty. In order to find a proper threshold, we create $10^5$ sets of variations of the observed intensities, $\{\{I_i\}_j\}$, so that the value of each intensity, $I_{i,j}$, comes from a normal distribution with mean $I_i$ and standard deviation $\Delta I_i$. Then, we calculate the loss for the optimised model with equation \ref{equation-loss} for each set of intensities $\{I_i\}_j$, obtaining a distribution of losses. We choose the threshold so that the loss values lower than such a threshold are the 68.27 percent of the total values (similarly to the definition of $1\,\upsigma$ confidence interval, but for an asymmetric distribution with a minimum value). The threshold thus corresponds to the 68.27 percentile (see Fig. \ref{figure-loss-hist}).

Then, we have to explore our parameter space around the minimised values in order to find when the loss function reaches our threshold value. If we only fit the column density, we only have to calculate the loss function in the surroundings of the minimised value until we reach the threshold value. However, if we also want to minimise the H$_2$ number density, we have to explore a two-dimensional parameter space. We opted to make an adaptive grid that starts calculating the loss values at the surroundings of the obtained minimum and continues enlarging its size until it encloses all the loss values minor to the threshold with a reasonable margin. This way, we obtained Figs. \ref{figure-loss-min-1} and \ref{figure-loss-min-2}.

\begin{figure}
\begin{center}
\includegraphics[trim={0 8 0 0}, clip, width=\hsize]{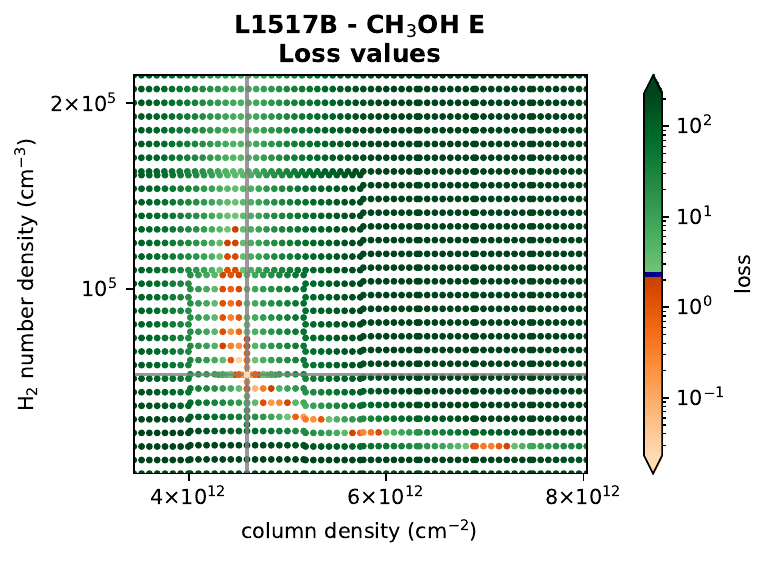}
\end{center}
\caption{Plot of the loss ($\mathcal{L}$) versus the column density ($N$) and the H$_2$ number density ($n_\mathrm{H_2}$) for a range of values around the optimised values that minimise the loss function, for the case of methanol E in the dust peak of L1517B. The gray lines indicate the optimised values for $N$ and $n_\mathrm{H_2}$. The color map is composed by two ranges of colors in order to highlight the points in which the loss ($\mathcal{L}$) is lower than the threshold value (marked in blue in the color bar), which will define the uncertainties for both optimised parameters.}
\label{figure-loss-min-1}
\end{figure}

\begin{figure}
\begin{center}
\includegraphics[trim={0 8 0 0}, clip, width=\hsize]{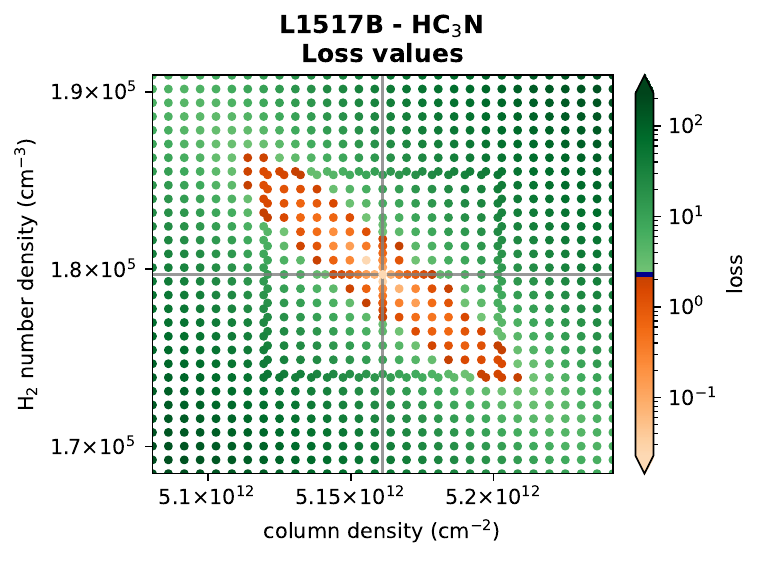}
\end{center}
\caption{Same as Fig. \ref{figure-loss-min-1}, but for cyanoacetilene (HC$_3$N) in the dust peak of L1517B.}
\label{figure-loss-min-2}
\end{figure}

Following this procedure, we have written a Python script that uses \textsc{Radex} and analyses our data in an automated way.\footnote{\url{https://github.com/andresmegias/radex-line-fitter/}} In this way, we obtained the column density values for methanol, acetonitrile and cyanoacetilene in L1517B shown in Table \ref{abunds-table}.

\section {Propagation of uncertainties and treatment of upper limits} \label{appendix-b}

Let's consider a group of $n$ variables $\{x_i\}$, with central values $\mu_i$ and uncertainties $\sigma_i$. This means that the probability density function (PDF) of the variable $x_i$ is centered around $\mu_i$ with a width of the order of $\sigma_i$, so that the $1\, \upsigma$ confidence interval (68.27 percent) is ($\mu_i - \sigma_i$, $\mu_i + \sigma_i$). To propagate the uncertainties through a function $f$ applied to the variables $\{x_i\}$, we can draw a sample of a large number of values $(\gtrsim\,10^4\sqrt{n})$ of each variable $x_i$, apply the function to each of the elements of the samples, and then obtain a central value and an uncertainty for the resulting distribution.

To do so, we need two things: an appropiate PDF for converting each variable $x_i$ to a distribution of values, and a proper method to obtain a central value and an uncertainty from the resulting distribution. For the last task, we can use the mean or the median as the central value,\footnote{We prefer to use the median, as it is more robust to outliers.} and the $1\, \upsigma$ confidence interval with respect to it (that includes 68.27 \% of the distribution) to obtain the lower and upper uncertainties. As for the PDF, if the domain of the variable $x$ is ($-\infty$, $\infty$), a proper function is the normal distribution:
\begin{equation}
    \indent
    \label{equation-normal-pdf}
    f(x) \;=\; \frac{1}{\sqrt{\uptau} \; \sigma} \exp \left (- \frac{1}{2} \left (\frac{x-\mu}{\sigma} \right )^2 \right ) \;\; ,
\end{equation}
with $\uptau \equiv 2\uppi$. However, if the domain of the variable is not ($-\infty$, $\infty$), this function would be incorrect. Therefore, we have to use another function as the PDF.

Let's suppose a domain ($b_1$, $b_2$). We define the left and right amplitudes, $a_1$ and $a_2$, as the distances between the limits of the domain and the median, that is, $a_j = |b_j - \mu|$ for $j = 1,2$. Now, as these amplitudes can be different, we will split our desired PDF in two halfs, one for $x \leq \mu$ and other for $x > \mu$. Then, we will use an amplitude $a$ as a reference, which must be greater than the uncertainty, $a > \sigma$. If the amplitude is quite greater than the uncertainty, $a \gg \sigma$, a good PDF would be just the normal distribution truncated to the domain ($b_1$, $b_2$). However, for amplitudes closer to the uncertainty, it would be considerably incorrect, as the truncation shifts the median of the distribution and modifies the confidence intervals, and thus the uncertainties. To fix this, we make the following variable change:
\begin{equation}
    \indent
    \label{equation-variable-change}
    x - \mu \;\; \rightarrow \;\; \tilde x - \mu \; \equiv \; \frac{4}{\uptau} \, a \, \tan \left ( \frac{\uptau}{4} \; \frac{x - \mu}{a} \right ) \,\, .
\end{equation}
Using this new variable $\tilde x$ with a normal distribution, we are able to compress the original domain of ($-\infty$, $\infty$) to ($-a$, $a$). However, we get two disadvantages: firstly, the normalization constant is now different, and secondly, the relation between the parameter of the standard deviation of the original normal distribution and the $1\, \upsigma$ confidence interval (from which we define the uncertainty, $\sigma$) is now different; therefore, we should rename the standard deviation of the original normal distribution to $s$, which we will call \textit{width}. The first change is not a problem, as the PDF will be normalised computationally for each case. As for the second one, we have characterised computationally the relation between $\sigma$ and $s$. It happens that $s / \sigma$ decreases with $a/\sigma$, having $s / \sigma \simeq 2.65$ when $a/\sigma$ = 2 and $s / \sigma \rightarrow 1$ when $a/\sigma \rightarrow \infty$. We have then a relation between the width, the uncertainty and the amplitude, that is, $s = s(\sigma, a)$.

Therefore, the resulting PDF would be the following:
\begin{equation}
    \indent
    \label{equation-general-normal-pdf}
    f(x) \; \propto \; \exp \left (- \frac{1}{2} \left (\frac{\frac{4}{\uptau}  a  \tan \left ( \frac{\uptau}{4} \frac{x - \mu}{a} \right)}{s(\sigma,a)} \right )^2 \right ) \;\; .
\end{equation}
We observed that $s / \sigma$ increases quasi-exponentially as $a/\sigma$ approaches a minimum value of $\sim\,$1.47. Moreover, we discovered that if $a/\sigma \lesssim 1.7$, the estimated relation $s(\sigma, a)$ starts to be incorrect because of the increasing dispersion in $s$. Therefore, we must use another PDF for the cases in which $a/\sigma \lesssim 1.7$. Actually, for that limit case the shape of the PDF is almost a uniform distribution between $-a$ and $a$. We can model this PDF as a trapezoidal function with a rectangular core and triangles in the edges. By doing so, one can easily demonstrate that this kind of PDF would only work if $a / \sigma$ is greater than the inverse of the integrated area corresponding to the $1\,\upsigma$ confidence interval ($\sim\,$0.683); that is, $a / \sigma \gtrsim1.46$, which is consistent with the found asymptote of the calculated relationship $s(\sigma, a)$.

\begin{figure}
\begin{center}
\includegraphics[trim={0 8 0 0}, clip, width=\hsize]{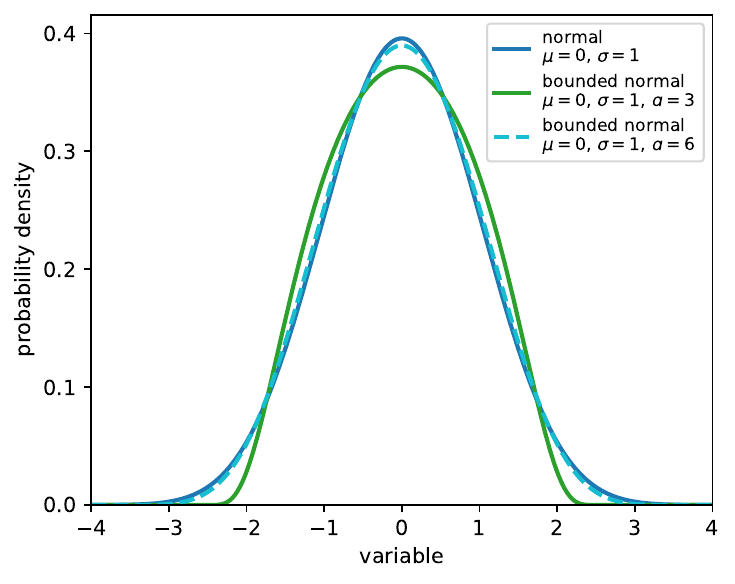}
\end{center}
\caption{Probability density functions for a normal distribution and two bounded normal distributions with diferent domains.}
\label{figure-bounded-gaussian}
\end{figure}

We then choose to only use this distribution when $a \geq 2 \sigma$, for the sake of simplicity and following a conservative approach. For $a \gg \sigma$, this PDF tends to a normal PDF, so we use this general PDF instead of a truncated gaussian for this regime (see Figure \ref{figure-bounded-gaussian}). We call this distribution a \textit{bounded normal distribution}.

Finally, for $\sigma < a < 2 \sigma$, we use a shifted and mirrored lognormal distribution whose PDF would be:
\begin{multline}
    \indent
    \label{equation-lognormal-pdf}
    f(x) \;=\; \frac{1}{\sqrt{\uptau} \; a \, \left | \mathrm{ln}\left( 1 - \frac{\sigma}{a} \right) \right |} \; \frac{1}{ 1 - \frac{|x-\mu|}{a}} \\ \times \;\, \exp \left (- \frac{1}{2} \left (\frac{\mathrm{ln}\left ( 1 - \frac{|x - \mu|}{a} \right )}{\mathrm{ln}\left ( 1 - \frac{\sigma}{a} \right )} \right )^2 \right ) \;\;\; .
\end{multline}
This PDF meets our requirements: the median is equal to $\mu$ and the uncertainty associated with the $1\, \upsigma$ confidence interval is $\sigma$. We call this distribution a \textit{mirrored lognormal distribution}.

We have defined our PDFs for the case of a variable $x$ with a central value $\mu$ and an uncertainty $\sigma$, building the final PDF with two halves with amplitude $a = a_j$ for $j = 1,2$. In case we had lower and upper uncertainties, $\sigma_1$ and $\sigma_2$, we should just replace $\sigma$ by $\sigma_1$ for the left half of the PDF and by $\sigma_2$ for the right half.

\begin{table*}
\label{table-abunds-ratio}

\begin{flushleft}
\small{\textbf{Table C1.} Column densities of HC$_3$N and CH$_{3}$CN ($N$) and abundance ratio
of this molecules ($\chi_{\mathrm{HC_3N}} \, / \, \chi_{\mathrm{CH_{3}CN}}$)
for the starless cores L1517B, L1498, L1544, and L1521E, 12 protostellar systems, 4 protoplanetary discs, and 2 comets.}
\end{flushleft}

\centering

\renewcommand*{\arraystretch}{1.5}

\begin{tabular}{ccccccc}
\toprule 
\multicolumn{4}{c}{\textbf{Source}} & $\boldsymbol{N_{\mathrm{HC_{3}N}}}\,(\mathrm{cm}{}^{-2})$ & $\boldsymbol{N_{\mathrm{CH_{3}CN}}}\,(\mathrm{cm}{}^{-2})$ & $\boldsymbol{\chi_{\mathrm{HC_{3}N}}\thinspace/\thinspace\chi_{\mathrm{CH_{3}CN}}}$\tabularnewline
\midrule
\multirow{7}{*}{{\small{}starless cores}} & \multirow{2}{*}{L1517B} & \multicolumn{2}{l}{{\footnotesize{}dust peak}} & $(5.16\thinspace\pm\thinspace0.05)\times10^{13}$ & $(2.1\thinspace\pm\thinspace0.6)\times10^{11}$ & $250_{-50}^{+100}$\tabularnewline
 &  & \multicolumn{2}{l}{{\footnotesize{}methanol peak}} & $3.72_{-0.13}^{+0.14}\times10^{12}$ & $<\thinspace2.0\times10^{11}$ & $>\thinspace17$\tabularnewline
 & \multirow{2}{*}{L1498} & \multicolumn{2}{l}{{\footnotesize{}dust peak}} & $(1.6\thinspace\pm\thinspace0.3)\times10^{13}$ & $<\thinspace8\times10^{10}$ & $>\thinspace100$\tabularnewline
 &  & \multicolumn{2}{l}{{\footnotesize{}methanol peak}} & $(2.2\thinspace\pm\thinspace1.0)\times10^{13}$ & $(1.0\thinspace\pm\thinspace0.1)\times10^{11}$ & $220_{-100}^{+100}$\tabularnewline
 & \multirow{2}{*}{L1544} & \multicolumn{2}{l}{{\footnotesize{}dust peak}} & $(1.0\thinspace\pm\thinspace0.3)\times10^{14}$ & $(1.5\thinspace\pm\thinspace0.2)\times10^{11}$ & $670_{-200}^{+230}$\tabularnewline
 &  & \multicolumn{2}{l}{{\footnotesize{}methanol peak}} & $(4.2\thinspace\pm\thinspace0.4)\times10^{13}$ & $<\thinspace9.1\times10^{10}$ & $>\thinspace310$\tabularnewline
 & L1521E & \multicolumn{2}{l}{{\footnotesize{}dust peak}} & $(8.4\thinspace\pm\thinspace2.5)\times10^{12}$ & $(4.8\thinspace\pm\thinspace1.4)\times10^{11}$ & $18_{-6}^{+9}$\tabularnewline
\midrule 
\multirow{12}{*}{{\small{}protostellar systems}} & \multicolumn{3}{c}{B1-a} & $(4.2\thinspace\pm\thinspace1.2)\times10^{12}$ & $(4.9\thinspace\pm\thinspace1.1)\times10^{11}$ & $9_{-3}^{+3}$\tabularnewline
 & \multicolumn{3}{c}{B1-c} & $(4\thinspace\pm\thinspace3)\times10^{12}$ & $(3.5\thinspace\pm\thinspace0.6)\times10^{11}$ & $12_{-9}^{+11}$\tabularnewline
 & \multicolumn{3}{c}{B5 IRS1} & $(3.2\thinspace\pm\thinspace1.2)\times10^{12}$ & $<\thinspace1.7\times10^{11}$ & $>\thinspace5$\tabularnewline
 & \multicolumn{3}{c}{IRAS 03235} & $(3.9\thinspace\pm\thinspace1.0)\times10^{12}$ & $(1.6\thinspace\pm\thinspace0.9)\times10^{11}$ & $25_{-11}^{+31}$\tabularnewline
 & \multicolumn{3}{c}{IRAS 03245} & $(4\thinspace\pm\thinspace3)\times10^{12}$ & $(1.9\thinspace\pm\thinspace1.5)\times10^{11}$ & $19_{-15}^{+69}$\tabularnewline
 & \multicolumn{3}{c}{IRAS 03271} & $(1.7\thinspace\pm\thinspace1.4)\times10^{12}$ & $(1.9\thinspace\pm\thinspace1.1)\times10^{11}$ & $9_{-7}^{+14}$\tabularnewline
 & \multicolumn{3}{c}{IRAS 23238} & $(3.2\thinspace\pm\thinspace2.6)\times10^{12}$ & $(1.4\thinspace\pm\thinspace0.3)\times10^{11}$ & $22_{-18}^{+21}$\tabularnewline
 & \multicolumn{3}{c}{L1014 IRS} & $(4.4\thinspace\pm\thinspace3.6)\times10^{11}$ & $<\thinspace6\times10^{10}$ & $>\thinspace0.7$\tabularnewline
 & \multicolumn{3}{c}{L1455 IRS3} & $(6\thinspace\pm\thinspace5)\times10^{11}$ & $<\thinspace9\times10^{10}$ & $>\thinspace0.7$\tabularnewline
 & \multicolumn{3}{c}{L1455 SMM1} & $(2.4\thinspace\pm\thinspace1.9)\times10^{12}$ & $<\thinspace1.1\times10^{11}$ & $>\thinspace2.2$\tabularnewline
 & \multicolumn{3}{c}{L1489 IRS} & $(3.5\thinspace\pm\thinspace2.4)\times10^{12}$ & $<\thinspace1.4\times10^{11}$ & $>\thinspace0.3$\tabularnewline
 & \multicolumn{3}{c}{SVS 4-5} & $(1.1\thinspace\pm\thinspace0.3)\times10^{13}$ & $(5.2\thinspace\pm\thinspace0.9)\times10^{11}$ & $21_{-6}^{+8}$\tabularnewline
\midrule 
\multirow{4}{*}{{\small{}protoplanetary discs}} & \multicolumn{3}{c}{GM Aur} & $1.9_{-0.4}^{+0.4}\times10^{13}$ & $2.1_{-0.1}^{+0.2}\times10^{12}$ & $8.8_{-1.9}^{+2.0}$\tabularnewline
 & \multicolumn{3}{c}{As 209} & $2.9_{-0.5}^{+0.5}\times10^{13}$ & $1.7_{-0.2}^{+0.2}\times10^{12}$ & $17_{-3}^{+4}$\tabularnewline
 & \multicolumn{3}{c}{HD 163296} & $7_{-2}^{+3}\times10^{13}$ & $2.3_{-0.2}^{+0.2}\times10^{12}$ & $32_{-9}^{+11}$\tabularnewline
 & \multicolumn{3}{c}{MWC 480} & $8_{-3}^{+4}\times10^{13}$ & $3.5_{-0.2}^{+0.2}\times10^{12}$ & $22_{-8}^{+11}$\tabularnewline
\midrule 
\multicolumn{4}{c}{\textbf{Source}} & $\boldsymbol{\boldsymbol{\chi_{\mathrm{HC_{3}N}}\thinspace/\thinspace\chi_{\mathrm{H_{2}O}}}}$ & $\boldsymbol{\boldsymbol{\chi_{\mathrm{CH_{3}CN}}\thinspace/\thinspace\chi_{\mathrm{H_{2}O}}}}$ & $\boldsymbol{\chi_{\mathrm{HC_{3}N}}\thinspace/\thinspace\chi_{\mathrm{CH_{3}CN}}}$\tabularnewline
\midrule
\multirow{2}{*}{{\small{}comets}} & \multicolumn{3}{c}{46P} & $<\thinspace3\times10^{-5}$ & $(1.7\thinspace\pm\thinspace0.01)\times10^{-4}$ & $<\thinspace0.21$\tabularnewline
 & \multicolumn{3}{c}{67P} & $4\times10^{-6}$ & $5.9\times10^{-5}$ & $0.068$\tabularnewline
\bottomrule
\end{tabular} 

\,

\begin{justify}
\footnotesize{References for each source: L1517B: this work; L1498: \citet{jimenez-serra-2021}; L1544: \citet{jimenez-serra-2016}; L1521E: \cite{nagy-2019}; protostellar systems: \citet{bergner-2017}; protoplanetary discs: \citet{ilee-2021}; comets: \citet{biver-2021}.}
\end{justify}

\end{table*}

\renewcommand{\arraystretch}{1}

Lastly, we should also address the case of a variable with an upper/lower limit or even a finite interval. Let's consider an interval ($x_1$, $x_2$), which may be finite or infinite. If it is finite, we choose a uniform distribution between $x_1$ and $x_2$ as the corresponding PDF. But if it is infinite, we choose a log-uniform distribution with finite thresholds for 0 and $\pm \infty$, which we set to $\pm 10^{-90}$ and $\pm 10^{90}$. For example, for an interval of ($-100$, $\infty$), we would build a sample $\{x_{-}\}$ from a uniform distribution between $-90$ and $2$ and a sample $\{x_{+}\}$ from a uniform distribution between $-90$ and $90$. Our final distribution would be the joining of the samples of $\{-10^{\{x_{-}\}}\}$ and $\{10^{\{x_{+}\}}\}$.

To sum up, if we have a set of variables with central values $\mu_i$ and uncertainties $\sigma_1$, $\sigma_2$, we first build distributions $\{x_i\}$ using the mentioned PDFs. Then, we apply the function to the distributions, $f(\{x_i\})$, obtaining a new distribution. Finally, we use an algorithm to detect if the distribution corresponds to an interval (that can be an upper/lower limit) or a defined value with uncertainties, and derive the corresponding parameters.

Using this approach, we have written a Python library that allows to deal with values with uncertainties and upper/lower limits, performing the uncertainty propagation automatically.\footnote{\url{https://pypi.org/project/richvalues/}} We have used it throughout the calculations of this paper, e.g., for obtaining the values shown in Fig. \ref{figure-abunds-ratio}.

\section {Abundance ratio of cianoacetylene and acetonitrile} \label{appendix-c}

Table \textcolor{blue}{C1} shows the abundance ratio of cyanoacetylene (HC$_3$N) and acetonitrile (CH$_3$CN) through different sources, from starless cores to comets. The propagation of the uncertainties and the upper limits in the column density is done through simulations using statistical distributions (see Appendix \ref{appendix-b}).


\bsp	
\label{lastpage}
\end{document}